\documentclass[%
 reprint,
 amsmath,amssymb,
 aps,
 nofootinbib,
]{revtex4-1}

\usepackage{graphicx}
\usepackage{dcolumn}
\usepackage{bm}
\usepackage{titlesec}
\usepackage{alphalph}
\usepackage{hyperref}
\setcitestyle{super}
\usepackage{siunitx}

\newsavebox{\foobox}
\newcommand{\slantbox}[2][0]{\mbox{%
        \sbox{\foobox}{#2}%
        \hskip\wd\foobox
        \pdfsave
        \pdfsetmatrix{1 0 #1 1}%
        \llap{\usebox{\foobox}}%
        \pdfrestore
}}
\newcommand\unslant[2][-.25]{\slantbox[#1]{$#2$}}

\DeclareSIUnit{\micrometer}{\unslant{\mu}m}
\DeclareSIUnit{\microgram}{\unslant{\mu}\gram}
\DeclareSIUnit{\microM}{\unslant{\mu}M}
\DeclareSIUnit{\microliter}{\unslant{\mu}\liter}

\begin{document}

\title{Verticalization of bacterial biofilms}

\author{Farzan Beroz, Jing Yan, Yigal Meir, Benedikt Sabass, Howard A. Stone, Bonnie L. Bassler, and Ned S. Wingreen*}

\begin{abstract} 
Biofilms are communities of bacteria adhered to surfaces. Recently, biofilms of rod-shaped bacteria were observed at single-cell resolution and shown to develop from a disordered, two-dimensional layer of founder cells into a three-dimensional structure with a vertically-aligned core. Here, we elucidate the physical mechanism underpinning this transition using a combination of agent-based and continuum modeling. We find that verticalization proceeds through a series of localized mechanical instabilities on the cellular scale. For short cells, these instabilities are primarily triggered by cell division, whereas long cells are more likely to be peeled off the surface by nearby vertical cells, creating an ``inverse domino effect''. The interplay between cell growth and cell verticalization gives rise to an exotic mechanical state in which the effective surface pressure becomes constant throughout the growing core of the biofilm surface layer. This dynamical isobaricity determines the expansion speed of a biofilm cluster and thereby governs how cells access the third dimension. In particular, theory predicts that a longer average cell length yields more rapidly expanding, flatter biofilms. We experimentally show that such changes in biofilm development occur by exploiting chemicals that modulate cell length.
\end{abstract}

\maketitle

\renewcommand{\figurename}{Figure}

\titleformat{\section}    
       {\normalfont\fontfamily{cmr}\fontsize{12}{17}\bfseries}{\thesection}{1em}{}
\titleformat{\subsection}[runin]
{\normalfont\fontfamily{cmr}\bfseries}{}{1em}{}

\renewcommand\thefigure{\arabic{figure}}    


\setcounter{section}{0}
\setcounter{figure}{0}
\setcounter{equation}{0}

\titlespacing*{\subsection}
{0pt}{3ex plus 1ex minus .2ex}{10ex plus .2ex}


Biofilms are groups of bacteria adhered to surfaces \cite{R22, R21, VG}. These bacterial communities are common in nature, and foster the survival and growth of their constituent cells. A deep understanding of biofilm structure and development promises important health and industrial applications \cite{R20, R23}. Unfortunately, little is known about the microstructural features of biofilms due to difficulties encountered in imaging individual cells inside large assemblies of densely-packed cells. Recently, however, advances in imaging technology have made it possible to observe growing, three-dimensional biofilms at single-cell resolution\cite{R16,R1, R4}.

In the case of \emph{Vibrio cholerae}, the rod-shaped bacterium responsible for the pandemic disease cholera\cite{R24, R5}, high-resolution imaging revealed a surprisingly complex biofilm developmental program\cite{R1, R4}. Over the course of 12-24 hours of growth, an individual founder cell gives rise to a dome-shaped biofilm cluster, containing thousands of cells that are strongly vertically ordered, especially at the cluster core. Notably, this ordering is an intrinsically non-equilibrium phenomenon, as it is driven by growth, not by thermal fluctuations. Indeed, the striking extent of ordering cannot be explained by Onsager's theory for the equilibrium ordering of rod-shaped objects\cite{R1,R31}.

An important clue to understanding the emergence of vertical order in \emph{V. cholerae} biofilms comes from genetic analyses that established the biological components relevant for biofilm development\cite{R25, R1, R4}. To facilitate their growth as biofilms, \emph{V. cholerae} cells secrete adhesive matrix components: Vibrio polysaccharide (VPS), a polymer that expands to fill gaps between cells, and cell-to-cell and cell-to-surface adhesion proteins. Cell-to-surface interactions enable vertical ordering by breaking overall rotational symmetry. However, despite previous work on the orientational dynamics of bacterial cells\cite{R6,R9,R27,AR1,AR2,R3,R17,R8,R7,GS}, the nature of this physical process remains unclear.

In this work, we establish the biophysical mechanisms controlling \emph{V. cholerae} biofilm development. We show that the observed structural and dynamical features of growing biofilms can be reproduced by a simple, agent-based model. Our model treats individual cells as growing and dividing rods with cell-to-cell and cell-to-surface interactions, and thus serves as a minimal model for a wide range of biofilm-forming bacterial species. By examining individual cell verticalization events, we show that reorientation is driven by localized mechanical instabilities occurring in regions of surface cells subject to high in-plane compression. These threshold instabilities explain the tendency of surface-adhered cells to reorient rapidly following cell division. We incorporate these verticalization instabilities into a continuum theory, which allows us to predict the expansion speed of biofilms as well as overall biofilm morphology as a function of cell-scale properties. We verify these predictions in experiments in which we use chemicals that alter cell length. Our model thus elucidates how the mechanical and geometrical features of individual cells control the emergent features of the biofilm, which are relevant to the survival of the collective.

\begin{figure*}[t!]
\includegraphics[width=1.6\columnwidth]{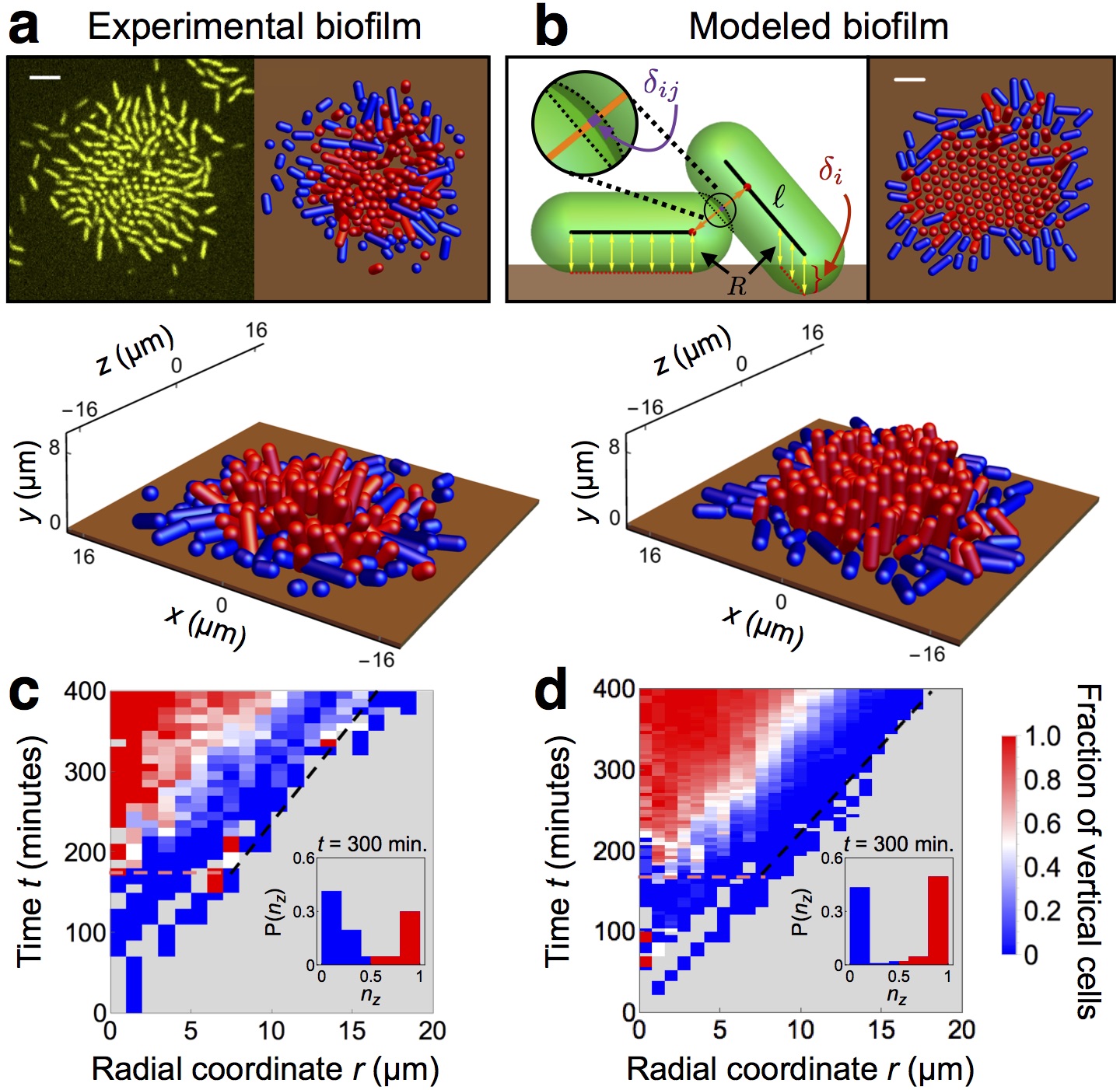}
\caption{\label{fig:FG0}
Development of experimental and modeled biofilms. (\textbf{a}, \textbf{b}) Top-down and perspective visualizations of the surface layer of (\textbf{a}) experimental and (\textbf{b}) modeled biofilms, showing positions and orientations of horizontal (blue) and vertical (red) surface-adhered cells as spherocylinders of radius $R=\SI{0.8}{\micrometer}$, with the surface shown at height $z=\SI{0}{\micrometer}$ (brown). Cells with $n_z < 0.5$ ($>0.5$) are considered horizontal (vertical), where $\boldsymbol{\hat{n}}$ is the orientation vector. The upper-left panel of (\textbf{a}) shows a confocal fluorescence microscopy image, and the upper-right panel shows the corresponding reconstructed central cluster using the positions and orientations of surface cells. The upper-left panel of (\textbf{b}) shows a schematic representation of modeled cell-cell (orange) and cell-surface (yellow) interactions, which depend, respectively, on the cell-cell overlap $\delta_{ij}$ (purple) and cell-surface overlap $\delta_i$ (red) (see Methods for details). Scale bars: $\SI{5}{\micrometer}$. (\textbf{c}, \textbf{d}) 2D growth of biofilm surface layer for (\textbf{c}) experimental biofilm (same as shown in (\textbf{a})) and (\textbf{d}) modeled biofilms. The color of each spatiotemporal bin indicates the fraction of vertical cells at a given radius from the biofilm center, averaged over the angular coordinates of the biofilm (gray regions contain no cells). In (\textbf{d}), each spatiotemporal bin is averaged over ten simulated biofilms. In (\textbf{c},\textbf{d}), the horizontal dashed pink lines show the onset of verticalization. The black dashed lines show the edge of the biofilm. Insets show the distribution of cell orientations at time $t=300$ minutes, with color highlighting horizontal and vertical orientations.
}
\end{figure*}

\subsection*{Biofilm radius and vertical ordering spread linearly over time} ~

How do cells in \emph{V. cholerae} biofilms become vertical? Biofilms grown from a single, surface-adhered founder cell initially expand along the surface (Fig. 1a, Supplementary Video 1). This horizontal expansion occurs because cells grow and divide along their long axes, which remain parallel to the surface due to cell-to-surface adhesion\cite{R1}. After about three hours, progeny near the biofilm center begin to reorient away from the surface (Fig. 1c). Reorientation events typically involve a sharp change in a cell's verticality $n_z$, defined as the component of the cell-orientation vector $\boldsymbol{\hat{n}}$ normal to the surface (inset Fig. 1c). At later times, the locations of the reorientation events spread outward, and eventually the biofilm develops a roughly circular region of vertical cells surrounded by an annular region of horizontal cells. Both of these regions subsequently expand outward with approximately equal, fixed velocities. The radial profile of verticality versus time shows that the local transition of cells from horizontal to vertical occurs rapidly, taking $10$-$30$ minutes for cell-sized regions to develop a vertical majority.

\subsection*{Agent-based model captures spreading of biofilm and vertical ordering}~

To understand how the behavior of living biofilms arises from local interactions, we developed an agent-based model for biofilm growth (left inset Fig. 1b, Methods, Supplementary Fig. 1). The model extends existing agent-based models\cite{R18, R15, R3} by incorporating the viscoelastic cell-to-surface adhesion\cite{R11,R19} that is crucial for \emph{V. cholerae} biofilm formation\cite{R24, R25}. Specifically, we treat the cells as soft spherocylinders that grow, divide, and adhere to the surface. We simulated biofilms by numerically integrating the equations of motion starting from a single, surface-adhered founder cell (Supplementary Video 2, Methods). To make the computations more tractable for systematic studies, we simulated only the surface layer of cells by removing from the simulation cells that become detached from the surface. This quasi-3D model is a reasonable approximation of the full 3D model at early times (Supplementary Fig. 2), and closely matches the dynamics and orientational order observed over the full duration of the experiment, including an inner region of vertical cells surrounded by an annular periphery of horizontal cells, both of which expand outward at a fixed rate (Fig. 1b,d). As in the experimental biofilm, the horizontal and vertical orientations of the modeled cells are sharply distinct (inset Fig. 1d), and the conversion from horizontal to vertical occurs rapidly (Fig. 1d). The emergence of this distinctive orientational-temporal pattern demonstrates that our simple agent-based model is sufficient to capture the physical interactions that underpin the observed ``verticalization'' transition of experimental biofilms.

\subsection*{Mechanical instabilities cause cell verticalization}~

Why do the experimental and modeled cells segregate into horizontal and vertical orientations, with transitions from horizontal to vertical proceeding rapidly? To investigate the local mechanics that drive verticalization, we considered the dynamics of a single model cell of cylinder length $\ell$ adhered to the surface. The surface provides a combination of attractive and repulsive forces that, in the absence of external forces, maintain the cell at a stable fixed point with elevation angle $\theta=0$ (i.e. horizontal) and penetration into the surface $\delta_0$. However, when additional forces are applied to the cell, the cell may become unstable to vertical reorientation.

We determined the onset of this instability by performing a linear stability analysis for a cell under constant external forces (inset Fig. 2a). For simplicity, we took the external forces to be applied by a continuum of rigid, spherical pistons that are distributed uniformly around the cell perimeter. The pistons compress the cell in the $xy$ plane with an applied surface pressure $p$. For values of $p$ larger than a threshold surface pressure $p_{t}$, the cell becomes linearly unstable to spontaneous reorientation (Supplementary Figs. 3-4). Our model yields a value of $p_{t}$ that increases with $\ell$. In particular, over a broad range of $\ell$, we find a simple linear increase of $p_{t}$ with $\ell$ (Fig. 2a). Intuitively, this increase occurs because the surface adhesion of the model cell scales with its contact area, creating an energy barrier to reorientation that increases with cell length.

To determine whether this simple model can predict verticalization events in the biofilm surface-layer simulations, we examined the forces acting on individual modeled cells throughout the development of a biofilm. Specifically, we computed the reorientation ``surface pressure'' $p_{r}$, defined as the sum of the magnitudes of the in-plane cell-cell contact forces on a cell, normalized by the perimeter of its footprint, at the instant it begins to reorient. We determined the instant of reorientation as the time of the peak of the total in-plane force on a cell immediately prior to it becoming vertical (Supplementary Fig. 5). We found that the average reorientation surface pressure $\langle p_{r} \rangle$ increases with $\ell$, as expected from the compressive instability model (Fig. 2a). Furthermore, the predicted value $p_t$ is in good agreement with the observed $\langle p_{r} \rangle$ for short cells. However, for long cells, $\langle p_{r} \rangle$ saturates more rapidly than $p_{t}$.

\begin{figure*}[t!]
\includegraphics[width=1.6\columnwidth]{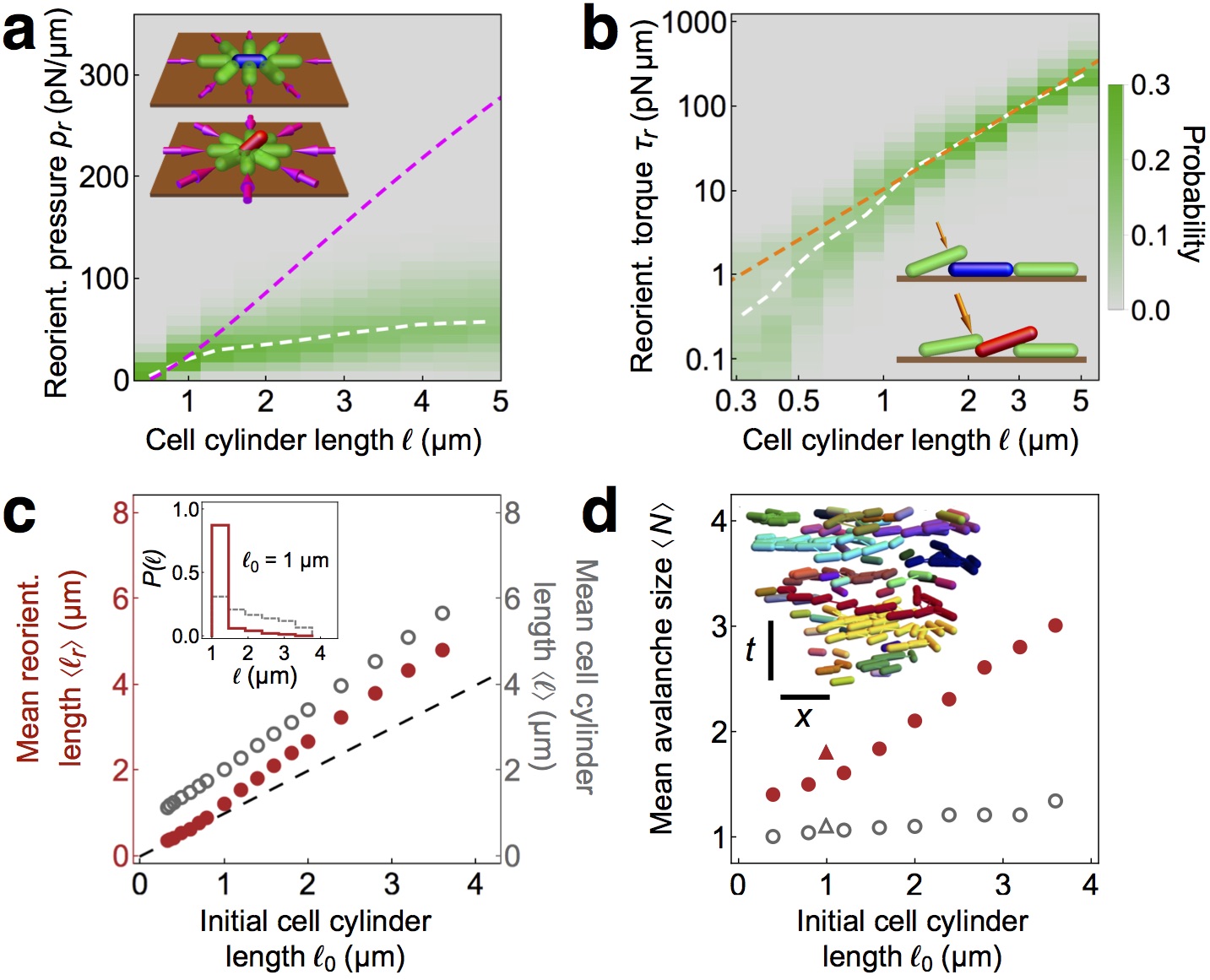}
\caption{\label{fig:FG0} Mechanics of cell reorientation in modeled biofilms. (\textbf{a}-\textbf{b}) Properties of individual cells at the time $t_{r}$ of reorientation, defined as the time of the peak of total force on the cell prior to it becoming vertical. Analyses are shown for all reorientation events among different biofilms simulated for a range of initial cell lengths $\ell_0$. (\textbf{a}) Distributions of reorientation ``surface pressure'' $p_{r}$, defined as the total contact force in the $xy$ plane acting on a cell at time $t_{r}$, normalized by the cell's perimeter, versus cell cylinder length $\ell$. The white dashed curve shows the average reorientation surface pressure $\langle p_{r} \rangle$ as a function of $\ell$. The magenta dashed curve shows the theoretical prediction for $\langle p_{r} \rangle$ from linear stability analysis for a modeled cell under uniform pressure, depicted schematically in the inset. (\textbf{b}) Distributions of the logarithm of reorientation torque $\tau_{r}$, defined as the magnitude of the torque on a cell due to cell-cell contact forces in the $z$ direction at time $t_{r}$, for different cell cylinder lengths $\ell$. The white dashed curve shows the average values $\langle \log \tau_{r} \rangle$ as a function of $\ell$. The orange dashed curve shows the scaling prediction $\tau_{r} \sim \ell^2$ from linear stability analysis for a modeled cell under torque, depicted schematically in the inset. (\textbf{c}) Mean reorientation length $\langle \ell_{r} \rangle$ (red), defined as the average value of cell length at $t_{r}$, and mean cell cylinder length $\langle \ell \rangle$ (gray), defined as the average length of all horizontal cells over all times of biofilm growth, averaged over ten simulated biofilms, each with initial cell cylinder length $\ell_0$, plotted versus $\ell_0$. The inset shows the distribution of reorientation lengths (red) and horizontal surface-cell lengths (gray) for $\ell_0 = \SI{1}{\micrometer}$. (\textbf{d}) Mean avalanche size $\langle N \rangle$, defined as the average size of a cluster of reorienting cells that are proximal in space and time (Supplementary Figs. 8-10), versus initial cell length $\ell_0$ for the experimental biofilm (red triangle) and the modeled biofilm (red circles). Open gray triangle and circles indicate the corresponding mean avalanche sizes for a null model. Inset shows a side view of cell configurations in the $xy$ plane at times $t_{r}$ for all reorientation events in a simulated biofilm with $\ell_0 = \SI{2.5}{\micrometer}$. Reorientation events are colored alike if they belong to the same avalanche. Scale bars: $\SI{10}{\micrometer}$ and $1$ hour.
}
\end{figure*}

\subsection*{The dominant mechanism of verticalization depends on cell length}~

How can longer cells become vertical at surface pressures much lower than the threshold values predicted by the compressive instability model? The large extent of the discrepancy suggests that for long cells, the in-plane forces alone are insufficient to cause the instabilities. Indeed, incorporating the numerically-observed distribution of in-plane forces acting on cells into the compressive instability model does not significantly improve the prediction for $\langle p_r \rangle$ (Supplementary Fig. 6). Thus, we hypothesized that in the case of long cells, forces acting in the $z$ direction might play an important role in triggering verticalization.

To explore this idea, we returned to the single-cell model and considered the effect of forces in the $z$ direction (inset Fig. 2b). Applying small forces in the $z$ direction to a cell with fixed center-of-mass height shifts the equilibrium elevation angle of the cell to a small finite value of $\theta$, proportional to the net torque. Under large enough torque, a single end of the cell becomes free of the surface, at which point the cell becomes unstable to further rotation and effectively ``peels'' off the surface. This nonlinearity, inherent in the geometry of contact, competes with the compressive instability, and under specific conditions can initiate reorientation at much smaller values of surface pressure. Specifically, for a fixed center-of-mass penetration depth $\delta_0 \ll \ell$, the threshold torque for peeling a cell off the surface due to forces in the $z$ direction scales as $\tau_t \sim \ell^2$ (Supplementary Fig. 4). For the whole-biofilm surface-layer simulation, the average reorientation torque, defined as the total torque due to forces in the $z$ direction at the instant of reorientation, closely obeys the predicted $\ell^2$ scaling for long cells (Fig. 2b). Taken together, our predictions from the compressive and peeling instabilities explain the verticalization of cells over the entire range of cell lengths studied.

\subsection*{Cell division can trigger verticalization}~

For both compressive and peeling instabilities, the presence of an energy barrier to reorientation explains the sharp distinction we observed between horizontal and vertical cell orientations. Furthermore, both mechanisms predict larger reorientation thresholds for longer cell lengths. Hence, the model suggests that shorter cells should reorient more readily. To confirm this effect in our simulated biofilms, we compared the distribution of reorientation lengths $\ell_{r}$, defined as the cell cylinder length at the instant of reorientation, to the full distribution of horizontal cell lengths for a series of simulations with different values of the initial cell cylinder length $\ell_0$ (Fig. 2c). For all values of $\ell_0$, we found that the mean reorientation cell length $\langle \ell_{r} \rangle$ is substantially smaller than the mean horizontal cell length. In addition, for simulations with average cell lengths comparable to those in our experiments, most reorientation events occur immediately after cell division. The limited time resolution of the experiments precludes a quantitative analysis of division-induced verticalization; nevertheless, the propensity for cell division to trigger reorientation is clearly observed in the experimental biofilm (Supplementary Video 1, Supplementary Fig. 7).

\begin{figure*}[t!]
\includegraphics[width=1.5\columnwidth]{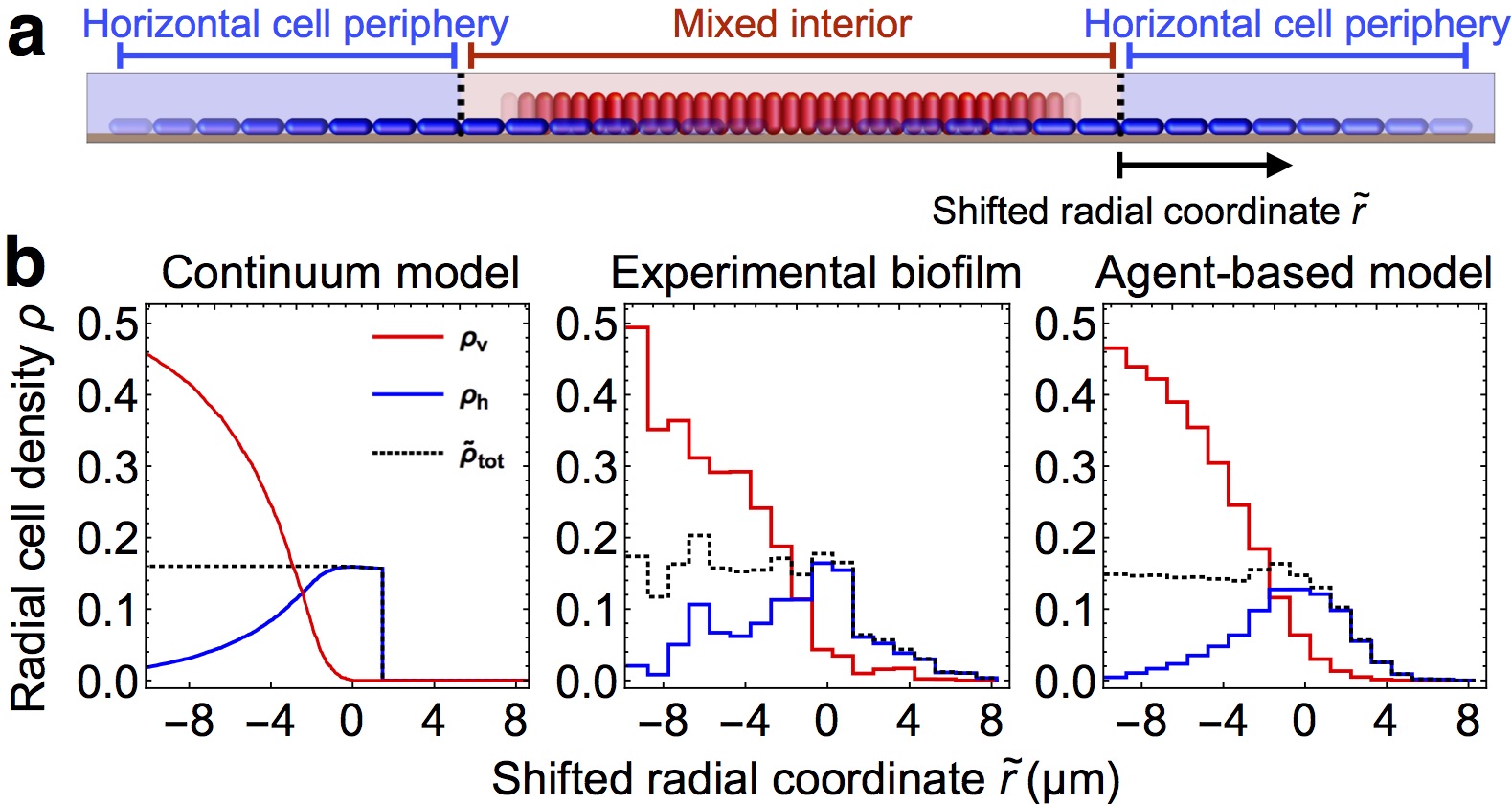}
\caption{\label{fig:FG0}
Two-component fluid model for verticalizing cells in biofilms. (\textbf{a}) Schematic illustration of the two-component continuum model. Horizontal cells (blue) and vertical cells (red) are modeled, respectively, by densities $\rho_h$ and $\rho_v$ in two spatial dimensions. The total cell density $\tilde{\rho}_{\mathrm{tot}}$ is defined as $\rho_h + \xi \rho_v$, where $\xi$ is the ratio of vertical to horizontal cell footprints. (\textbf{b}) Radial densities $\rho$ of vertical cells ($\rho_v$, red), horizontal cells ($\rho_h$, blue), and total density ($\tilde{\rho}_{\mathrm{tot}}$, black), versus shifted radial coordinate $\tilde{r}$, defined as the radial position relative to the boundary between the mixed interior and the horizontal cell periphery. Results are shown for the continuum model (left; radial cell density in units of $\SI{}{\micrometer}^{-2}$), the experimental biofilm (middle; radial cell density in each $\SI{}{\micrometer}$-sized bin averaged over an observation window of 50 minutes), and the agent-based model biofilm (right; radial cell density in each $\SI{}{\micrometer}$-sized bin averaged for ten biofilms over an observation window of $6$ minutes). For the continuum model and the agent-based model biofilms the parameters were chosen to match those obtained from the experiment (Supplementary Figs. 12-13).
}
\end{figure*}

\subsection*{Verticalization is localized}~

We next investigated how the surface compression and peeling instabilities influence the propagation of reorientation through a biofilm. First, we generalized our model for the surface compression instability to the multi-cell level. A linear stability analysis of the model suggests that reorientation events should be independent and spatially localized for short cell lengths (Supplementary Fig. 3). By contrast, for long cell lengths, the tendency of neighboring vertical cells to trigger reorientation suggests an (inverse) domino-like effect in which one cell standing up can induce its neighbors to stand up.

To quantify the extent of cooperative verticalization, we computed the size of reorientation ``avalanches'', defined as groups of verticalization events that are proximal in space and time\cite{ava} (inset Fig. 2d, Supplementary Fig. 8). We found that the mean avalanche size increases with cell length, consistent with the prediction of the inverse domino effect for long cells (Fig. 2d). Interestingly, however, the distribution of avalanche sizes decays roughly exponentially for all values of cell length we studied (Supplementary Fig. 9), with only a modest number of cells ($N \sim 1-3$) involved in typical avalanches (Fig. 2d). Our results indicate that the sizes of reorientation avalanches are limited by an emergent spatiotemporal scale governed by cell geometrical and mechanical properties, rather than by the growing total supply of horizontal cells.

A natural explanation for the small mean avalanche size comes from the reduction in cell footprint that occurs upon reorientation, which rapidly alleviates the local surface pressure responsible for verticalization (Supplementary Fig. 5). This effect combines with the disorder of the contact geometries and forces throughout the biofilm, which separates horizontal cells near the verticalization threshold into disconnected groups (Supplementary Video 3, Supplementary Fig. 10). Thus, although the inverse domino effect transiently increases verticalization cooperativity, avalanches quickly exhaust the supply of nearby horizontal cells that are susceptible to becoming vertical. Consequently, verticalization occurs throughout the biofilm in scattered, localized regions.

\subsection*{Two-fluid model describes propagation of verticalization}~

To understand how localized cell verticalization gives rise to the global patterning dynamics of the biofilm, we developed a two-dimensional continuum model that treats horizontal and vertical cell densities as two coupled fluids (Fig. 3a, Methods). The local horizontal cell density $\rho_h$ grows in the plane at a rate $\alpha$ and converts to vertical cell density $\rho_v$ in regions of high surface pressure or, equivalently in our model, high total 2D cell density $\tilde{\rho}_{\mathrm{tot}} \equiv \rho_h + \xi \rho_v$, where $\xi$ is the ratio of vertical to horizontal cell footprints. These interactions yield the following equation for the change in $\tilde{\rho}_{\mathrm{tot}}$ in regions of nonzero surface pressure:
\setlength{\abovedisplayskip}{12pt}
\setlength{\belowdisplayskip}{12pt}
\begin{equation}
\dot{\tilde{\rho}}_{\mathrm{tot}} = \gamma \nabla^2 \tilde{\rho}_{\mathrm{tot}} + \alpha  \rho_h - (1 - \xi) \beta \Theta(\tilde{\rho}_{\mathrm{tot}} - \tilde{\rho}_t) \rho_h,
\end{equation}

\noindent where $\gamma$ is the ratio of the Young's modulus of the biofilm to the surface drag coefficient, $\Theta$ is the Heaviside step function, and $\tilde{\rho}_t$ is the threshold surface density for verticalization. We simulated this continuum model, and found that for $\alpha < \beta$, the biofilm generically develops into a circular region containing both horizontal and vertical cells (``Mixed interior'') surrounded by an annular region containing horizontal cells (``Horizontal cell periphery''), closely matching both the experimental biofilm and the agent-based model biofilms (Fig. 3b, Supplementary Video 4, Supplementary Note). In this regime, the biofilm front spreads linearly in time at a fixed expansion speed $c^*$. Furthermore, the total cell density and the surface pressure are constant in the mixed interior. This constancy is stabilized by the competing effects of cell growth and cell verticalization, and occurs provided that $\alpha < \beta (1 - \xi)$ (Supplementary Fig. 11). When this condition is satisfied, verticalization can reduce cell density faster than cell density can be replenished by cell growth and cell transport due to gradients in surface pressure. Physically, this results in the cell density rapidly fluctuating around the verticalization threshold. This rapid alternation effectively tunes the verticalization rate down to $\alpha = \beta (1 - \xi)$, and thereby ensures a constant total cell density and surface pressure in the mixed interior. The resulting ``dynamical isobaricity'' (constancy of pressure) provides the boundary condition for the horizontal cell periphery that determines the horizontal expansion speed $c^*$, independent of $\beta$ and $\xi$. In the limit of slow expansion $c^* \ll \sqrt { \alpha \gamma } $, $c^*$ is given by:
\setlength{\abovedisplayskip}{12pt}
\setlength{\belowdisplayskip}{12pt}
\begin{equation}
c^* \simeq c^*_0 \sqrt{  1 - \frac{\tilde{\rho}_0}{\tilde{\rho}_{t}}  },
\end{equation}
\noindent where $c^*_0 = \sqrt{2 \alpha \gamma}$ and $\tilde{\rho}_0$ is the close-packed, but uncompressed, cell density. Thus, we find that $c^*$ increases with $\tilde{\rho}_{t}$ until it saturates to a maximum speed $c^*_0$ for $\tilde{\rho}_{t} \gg \tilde{\rho}_0$. Intuitively, higher values of the verticalization threshold density $\tilde{\rho}_{t}$ sustain a wider periphery of horizontal cells, which results in a higher rate of increase in the total number of surface cells. Thus, our continuum model reveals how the geometrical and mechanical properties of individual cells influence the global morphology of the growing biofilm.

\begin{figure}[t!]
\includegraphics[width=1\columnwidth]{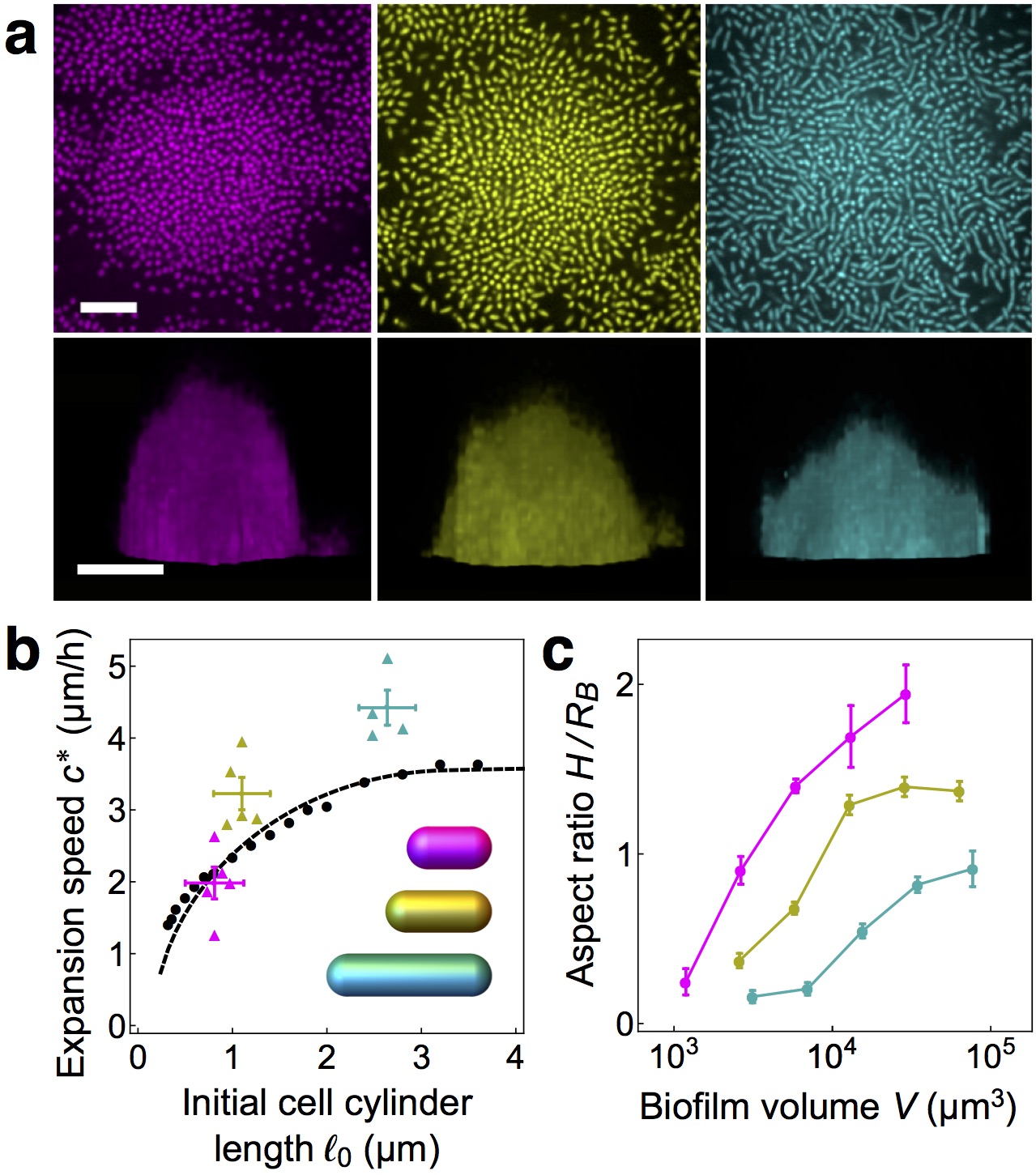}
\caption{\label{fig:FG0}
Global morphological properties of experimental and modeled biofilms. (\textbf{a}) Top-down (upper row) and side views (lower row) of experimental biofilms grown with $\SI{0.4}{\microgram/\milli\liter}$ A22 (magenta), without treatment (yellow), and with $\SI{4}{\microgram/\milli\liter}$ Cefalexin (cyan), following overnight growth (upper row) and 7 hours after inoculation (lower row). Scale bar: $\SI{10}{\micrometer}$. (\textbf{b}) Expansion speed $c^*$, defined as the speed of the biofilm edge along the surface, versus the initial cell cylinder length $\ell_0$ for experimental biofilms (A22, magenta; no treatment, yellow; Cefalexin, cyan), agent-based model biofilms (black circles), and continuum model (dashed black curve). Expansion velocities were determined from a linear fit of the basal radius $R_{B}$ of the biofilm versus time, where $R_{B}$ is defined at each time point as the radius of a circle with area equal to that of the biofilm base. For experimental biofilms, the boundary was extracted from the normalized fluorescence data (see Methods for details). For each treatment, the vertical error bars show the standard error of the mean of the expansion speed and the horizontal error bars bound the measured initial cell cylinder length (Supplementary Fig. 1). Inset: model cells with lengths and radii corresponding to the averages for different treatments. (\textbf{c}) Biofilm aspect ratio $H/R_{B}$ for experimental biofilms grown under different treatments, where the biofilm height is defined as $H=3V/2R_{B}^2$, the height of a semi-ellipsoid with a circular base of radius $R_{B}$ and volume $V$ equal to that of the biofilm. Color designations and treatments same as in panel (\textbf{a}). 
}
\end{figure}

\subsection*{Increasing cell length yields more rapidly expanding, flatter biofilms}~

Because the threshold surface density for verticalization $\tilde{\rho}_{t}$ increases with cell length, we expect biofilms composed of longer cells to maintain a wider periphery of horizontal cells and to thereby expand faster along the surface than biofilms composed of shorter cells. To test this notion in our agent-based model, we computed the expansion speed of the modeled biofilms for a range of initial cell cylinder lengths $\ell_0$ (Fig. 4b). Upon fitting the continuum model parameters to those of the agent-based model (Supplementary Figs. 12-13), we found that the expansion velocities of the two models were equal to within a few percent. In living, experimental biofilms, we can increase or decrease the average cell length using chemical treatments\cite{R5,T7} (Fig. 4a, top row). Similar to the agent-based model biofilms, in experimental biofilms, the surface expansion speed increases with increasing cell length (Fig. 4b, Supplementary Video 5). The experimentally observed speed appears to saturate as cell length is increased, as occurs in the modeled biofilms. Furthermore, when the model biofilm parameters are fitted to experiment (Supplementary Fig. 1), the experimental and model biofilm speeds agree to within twenty percent or better. Taken together, these observations support the conclusion that self-organized dynamical isobaricity governs the observed expansion of \emph{V. cholerae} biofilms.

How do different surface expansion speeds influence the ensuing biofilm development into the $z$ direction? After a few hours, living biofilms grow into roughly semi-ellipsoidal shapes with volume $V=(2/3)R_{B}^2 H$, where $R_{B}$ is the basal radius and $H$ is the height. For equal rates of total volume growth, we expect a biofilm that expands more rapidly along the surface to develop a lower aspect ratio $H/R_{B}$ than a biofilm that expands less rapidly along the surface. We verified that this trend holds for the experimental biofilms (Fig. 4a, bottom row). In particular, the measured aspect ratio $H/R_{B}$ increases with cell length over a wide range of volumes (Fig. 4c). Thus, our results show how the elongated geometries of individual cells govern the global morphology of the collective.

\subsection*{Discussion}~

Bacterial biofilms are pervasive lifeforms that significantly influence health and industry\cite{R22, R21, R20, R23,F1,KKLC}. An important step towards control over biofilms was achieved when the molecular building blocks of \emph{V. cholerae} biofilms were identified\cite{R25}. In particular, cell-to-surface adhesion factors were found to be necessary to generate vertically-ordered biofilm clusters\cite{R4}. Despite this progress, the dynamical process by which cells in biofilms become vertical has remained mysterious. Here, we showed that cell verticalization begins to occur when the local effective surface pressures that arise from cell growth become large enough to overcome the cell-to-surface adhesion that otherwise favors a horizontal orientation. Subsequently, the reduction in cell footprint that occurs upon cell verticalization, which acts to reduce the effective surface pressure, provides a mechanical feedback that controls the rate of biofilm expansion. Our continuum and agent-based models quantitatively capture the rate of horizontal expansion of experimental biofilms, and also predict the observed changes in the height-to-radius aspect ratio that occur with varying average cell length. \\
\indent Our results suggest that bacteria have harnessed the physics of mechanical instabilities to enable the generation of complex architectures. We expect that individual cell parameters have evolved in response to selective pressures on global biofilm morphology, e.g. during resource competition\cite{R12, R13,R16,T1,KCH}. Since optimal morphology may be condition dependent, cells may also have evolved adaptive strategies that alter biofilm architecture, which could be investigated experimentally by screening for environmental influences on cell size, shape, and surface adhesion\cite{T9}. \\
\indent For simplicity, we focused on flat surfaces, nutrient-rich conditions, and \emph{V. cholerae} strains that have been engineered to have simpler interactions than those in wild type biofilms (Methods). Moreover, our agent-based model does not explicitly incorporate the VPS matrix secreted by cells\cite{R21, osm, R15}. Understanding the modifying effects of the VPS matrix, cell and surface curvature (Supplementary Fig. 14), cell-to-cell adhesion (Supplementary Fig. 15), and chemical feedback\cite{T2} will be important directions for future studies. More broadly, we must develop a systematic method to account for the diversity of architectures that can be produced by local mechanical interactions (Supplementary Discussion). \\
\indent Our study of a two-fluid model for verticalizing biofilms led us to discover a novel type of front propagation. Interestingly, in the biofilm surface layer, the front profile of cell density is precisely uniform starting at some finite distance from the edge, whereas previous models of front propagation saturate asymptotically toward uniformity \cite{T4, T5, T6, BS}. The self-organized nature of this process yields a universal dependence of the expansion speed on the cell geometrical and mechanical parameters that is robust to details of the mechanical feedback. We have focused on the mean-field behavior of biofilms, but an open question is to understand the role of fluctuations in the ``pressure'' acting on cells, e.g. either from a jamming perspective\cite{R29}, a fluctuating hydrodynamical perspective\cite{T3, MLM}, or a combination of approaches. \\
\indent In summary, we have elucidated the physical mechanism underlying a complex developmental program observed at the cellular scale in bacterial biofilms. The relative biochemical and biophysical simplicity of this prokaryotic system allowed us to quantitatively understand the developmental pathway from the scale of a single cell to the scale of a large community assembly. Going forward, we expect bacterial biofilms will take on increasingly important roles as tractable models that can be used to understand how living systems generate and maintain their structures.

\begin{acknowledgments}
We thank Benjamin Bratton, Xiaowen Chen, Siddhartha Jena, Nishant Pappireddi, Pierre Ronceray, and Josh Shaevitz for insightful discussions and encouragement, Gary Laevsky for assistance with imaging, and Thomas Bartlett and Zemer Gitai for sharing the $\Delta crvA$ strain and reagents. This work was supported by the NIH under Grant No. R01 GM082938. (F. B., Y. M., $\&$ N. S. W.), the Howard Hughes Medical Institute (B. L. B.), NSF Grant MCB-0948112 (B. L. B.), NSF Grant MCB-1344191 (H. A. S., B. L. B., $\&$ N. S. W.), NIH Grant 2R37GM065859 (F. B. $\&$ B. L. B.), the Max Planck Society-Alexander von Humboldt Foundation (B. L. B.), and the Eric and Wendy Schmidt Transformative Technology Fund from Princeton University (H.A.S., B.L.B., $\&$ N.S.W.). J. Y. holds a Career Award at the Scientific Interface from the Burroughs Wellcome Fund.
\end{acknowledgments}

\section*{Affiliations}

\textbf{Joseph Henry Laboratories of Physics, Princeton University, Princeton NJ 08544, USA} \\
Farzan Beroz \\

\textbf{Department of Mechanical and Aerospace Engineering, Princeton University, NJ 08544, USA} \\
Jing Yan, Benedikt Sabass, Howard A. Stone \\

\textbf{Department of Molecular Biology, Princeton University, Princeton NJ 08544, USA} \\
Jing Yan, Bonnie L. Bassler, Ned S. Wingreen \\

\textbf{Department of Physics, Ben-Gurion University of the Negev, Beer-Sheva, Israel} \\
Yigal Meir \\

\textbf{ICS-2/IAS-2, Forschungszentrum J\"ulich, J\"ulich, Germany.} \\
Benedikt Sabass \\

\textbf{Lewis-Sigler Institute for Integrative Genomics, Princeton University, Princeton NJ 08544, USA}
Ned S. Wingreen

\section*{Author contributions}
F. B., J. Y., and N. S. W. conceived the research. F. B. performed all simulations and analysis. J.Y. carried out all experiments, and J. Y. and F. B. contributed to the experimental design. All authors contributed to the manuscript preparation.

\section*{Competing interests}
The authors declare no competing financial interests.

\section*{Methods}

\subsection*{Growing and imaging experimental biofilms} ~\\

\textbf{Strains and media.} The \emph{V. cholerae} strain used in this study is a derivative of wild-type \emph{Vibrio cholerae} O1 biovar El Tor strain C6706, harboring a missense mutation in the \emph{vpvC} gene (VpvC W240R) that elevates c-di-GMP levels and confers a rugose biofilm phenotype\cite{R26}. Additional mutations were engineered into this strain using \emph{Escherichia coli} S17-$\lambda$\emph{pir} carrying pKAS32. Specifically, the biofilm gene responsible for cell-cell adhesion, \emph{rbmA}, was deleted. To avoid the effects of cell curvature, we deleted \emph{crvA} encoding the periplasmic protein CrvA responsible for the curvature of \emph{V. cholerae} cells. Biofilm experiments were performed in M9 minimal medium, supplemented with 2 mM MgSO$_4$, $\SI{100}{\microM}$ CaCl$_2$, and 0.5$\%$ glucose. When indicated, Cefalexin (Sigma Aldrich) was added at $\SI{4}{\microgram/\milli\liter}$ and A22 (a gift from the Gitai group) was used at $\SI{0.4}{\microgram/\milli\liter}$. These concentrations were experimentally determined to modulate cell morphology without affecting overall mass accumulation. \\

\textbf{Biofilm growth.} \emph{V. cholerae} strains were grown overnight at 37 $^{\circ}$C in liquid LB medium with shaking, back-diluted 30-fold, and grown for an additional two hours with shaking in M9 medium until early exponential phase (OD$_{600}$ = $0.1 - 0.2$). These re-grown cultures were diluted to OD$_{600}$ = 0.001 and $\SI{100}{\microliter}$ of the diluted cultures were added to wells of 96-well plates with $\#$1.5 coverslip bottoms (MatTek). The cells were allowed to attach for ten minutes, after which the wells were washed twice with fresh M9 medium, and, subsequently, $\SI{100}{\microliter}$ of fresh M9 medium was added, with or without drugs. The low initial inoculation density enabled isolated biofilm clusters to form. The locations of the founder cells were identified, and one hour after inoculation, imaging was begun on the microscope stage at 25 $^{\circ}$C. \\

\textbf{Microscopy.} Details of the imaging system have been described elsewhere\cite{R4}. Briefly, images were acquired with a spinning disk confocal microscope (Yokogawa) a 543 nm laser (OEM DPSS), and an Andor iXon 897 EMCCD camera. For single-cell resolution imaging, a 60x water objective with a numerical aperture of 1.2 plus a 1.5x post-magnification lens was used. To avoid evaporation, immersion oil with a refractive index of 1.3300 $\pm$ 0.0002 (Cargille) was used instead of water. The time difference between each image acquisition was 10 minutes, and the total imaging time was 8 hours. Only the bottom $\SI{5}{\micrometer}$ of the biofilm was imagined (with a $z$ step size of $\SI{0.2}{\micrometer}$) to avoid excessive photobleaching and phototoxicity. For coarse-grained imaging, a 20x multi-immersion objective was used without post-magnification. In this case, the time difference between each image acquisition was 30 minutes, and entire biofilms were imaged (with a $z$ step size of $\SI{1}{\micrometer}$). At the end of the coarse-grained time course, the biofilm clusters were imaged again with high magnification to determine cell lengths. All image acquisitions were automated using Nikon Element software. All cells harbored mKO fluorescent proteins expressed from the chromosome. Experimental images in Fig. 4 were false-colored to differentiate between different growth conditions. \\

\textbf{Image processing.} The cell segmentation protocol and Matlab codes have been described elsewhere in detail\cite{R4}. Cell position, cell length, and cell orientation were used as input for 3D rendering in Fig. 1\textbf{a} and for further analysis. For biofilm clusters grown in the presence of A22 or Cefalexin, cell lengths were manually measured in the bottom cell layers of the biofilms using the Nikon Element software. To define the biofilm shape parameters in the coarse-grained images, the bottom cell layers of the biofilms were first identified by finding the brightest $z$-cross section, according to the total fluorescence intensity. The contour of the individual biofilm cluster was next identified using the three-dimensional Canny edge detection method implemented in Mathematica. To correct for the inevitable optical stretching in the $z$-direction, we compared the heights obtained from the same cluster in the coarse-grained and the fine resolution images. By imaging a series of biofilm clusters of different sizes, we obtained a curve of a biofilm cluster's actual height versus its apparent height in the coarse-grained images, which we used to calibrate the heights measured in the coarse-grained images. \\

\subsection*{Modeling biofilms} ~\\

\textbf{Agent-based model.} We model the volume occupied by a cell as a cylinder of length $\ell$ with two hemispherical end caps each of radius $R$. We treat cell growth as elongation that increases cell volume at a fixed rate $\alpha$ (chosen randomly at birth from a narrow Gaussian distribution to desynchronize cell divisions). Cells are born with an initial cell cylinder length $\ell_0$ and grow to twice their total length. When a cell reaches the division length, the cell is instantaneously replaced by two identical daughter cells. In our model, cell-to-cell and cell-to-surface overlaps exert repulsive forces according to Hertzian contact mechanics for elastic materials\cite{R32}. In the case of the cell-to-surface overlap, we also include an attractive interaction with an energy proportional to the cell-to-surface contact area, i.e., the Derjaguin approximation\cite{R11, R19}. Finally, we include two sources of viscous drag: a modest damping of three-dimensional motion through the surrounding fluid and biopolymer matrix, and a much larger damping of sliding motion along the adhesive surface\cite{R34}. We determine the parameters in our model by fitting to the experimental data (Supplementary Fig. 1). \\

\textbf{Simulation of agent-based model.} To simulate the agent-based model, we numerically integrate the equations of motion using an explicit embedded Runge-Kutta-Fehlberg method. Our implementation of this method in C++ is adapted from the \emph{GNU Scientific Library}\cite{M1}, and available freely online at GitHub\cite{M2}. To account for symmetry-breaking microscopic irregularities, we add a small amount of random noise to each component of the generalized force at every time step ($10^{-8}\ E_0 R^2$ to the force acting on the center-of-mass $\boldsymbol{r}$ and $10^{-8}\ E_0 R^3$ to the generalized force acting on the orientation vector $\boldsymbol{\hat{n}}$). The initial conditions consist of a single cell of length $\ell_0$ at elevation angle $\theta=0$ and penetration depth $\delta_0$. \\

\textbf{Continuum model.} To describe the radial expansion and orientational dynamics of the biofilm, we treated the horizontal cells and vertical cells as continuous fields $\rho_h$ and $\rho_v$, with the total density $\tilde{\rho}_{\mathrm{tot}}$ defined as the sum $\rho_h + \xi \rho_v$, where $\xi$ is the ratio of vertical to horizontal cell footprints. In regions where $\tilde{\rho}_{\mathrm{tot}}$ exceeds the close-packing density $\tilde{\rho}_0$, the rate of change of the horizontal cell density is proportional to the divergence of the flux of cells due to cell transport plus terms due to cell growth and cell verticalization. We assume that cell transport is given by the gradient of surface pressure divided by a surface viscosity coefficient $\eta_1$, where surface pressure is approximated as linearly proportional to the areal deformation $\tilde{\rho}_{\mathrm{tot}} - \tilde{\rho}_0$, i.e. following ``Hooke's law''. Cell growth in the plane is proportional to the local horizontal cell density and occurs at a fixed rate $\alpha$. We assume that cell verticalization locally converts horizontal cell density to vertical cell density at a fixed rate $\beta$ in regions where $\tilde{\rho}_{\mathrm{tot}} > \tilde{\rho}_{t}$, where $ \tilde{\rho}_{t}$ is the threshold surface density for verticalization. \\

\textbf{Simulation of continuum model.} We simulated the dynamics of the cell densities in Python using FiPy, a finite volume PDE solver\cite{M3}. We performed our simulation on a mesh containing $30000$ points with a spacing of $\Delta x=\SI{1}{\nano\meter}$ and a temporal step size of $\Delta t=\SI{1}{\milli\second}$. \\

\subsection*{Data availability} ~\\

The simulation used for the agent-based model is available on GitHub\cite{M2}. The code used for the analysis in the current study is available from the corresponding author following request.

\end{document}


\title{Verticalization of bacterial biofilms - Supplementary Information}

\author{Farzan Beroz, Jing Yan, Yigal Meir, Benedikt Sabass, Howard A. Stone, Bonnie L. Bassler, and Ned S. Wingreen$^*$}

\maketitle

\renewcommand{\figurename}{Supplementary Figure}
\renewcommand{\theequation}{S\arabic{equation}}

\titleformat{\section}    
       {\normalfont\fontfamily{cmr}\fontsize{12}{17}\bfseries\filcenter}{\thesection}{1em}{}
\titleformat{\subsection}[runin]
{\normalfont\fontfamily{cmr}\bfseries}{}{1em}{}

\renewcommand\thefigure{\arabic{figure}}    
\renewcommand\thesection{Supplementary Figure \arabic{section}:}

\onecolumngrid

\setcounter{section}{0}
\setcounter{figure}{0}
\setcounter{equation}{0}


\section{\textbf{Schematic illustration of agent-based model}}

\begin{figure}[H]
\centering
\includegraphics[width=0.8\columnwidth]{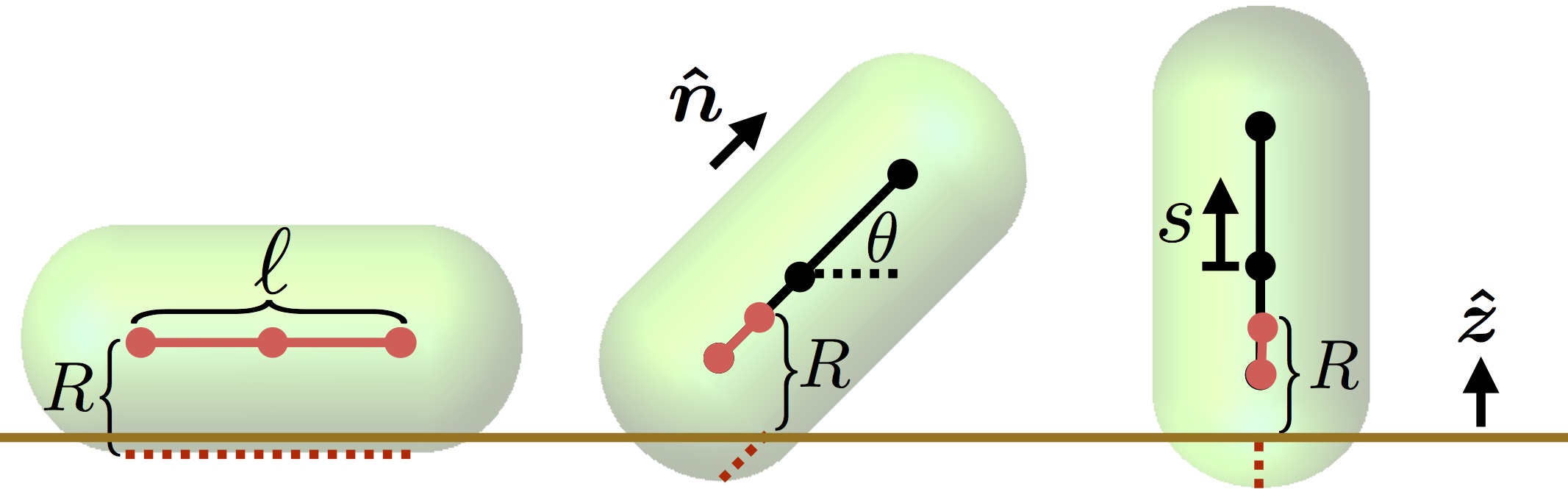}
\caption{\label{fig:FG0}
Cell model. Schematic of model cells, showing a horizontal cell (elevation angle $\theta = 0$, left), a cell that is angled with respect to the surface (middle), and a cell that is vertical ($\theta = \pi/2$, right). Cells are modeled as cylinders of length $\ell$ with two hemispherical endcaps of radius $R$. The cell orientation is specified by the unit vector $\boldsymbol{\hat{n}}$. The direction normal to the surface is specified by the unit vector $\boldsymbol{\hat{z}}$. The distance along the cell cylinder is parameterized by the coordinate $s$, which is zero at the cell's center of mass.
}
\end{figure}

\subsection*{Cell model}

We model each cell as a cylinder of length $\ell$ with hemispherical endcaps of radius $R$ (``spherocylinders'', Supplementary Fig. 1). In total, the volume $V$ of a model cell is therefore:

\begin{equation}
V = \frac{4}{3}\pi R^3 + \pi R^2 \ell.
\end{equation}

\noindent We treat cell growth as an increase in $\ell$ at a fixed radius $R$ with total volume growing at a rate $\alpha$:

\begin{equation}
\frac{dV}{dt} = \alpha V.
\end{equation}

\noindent Thus, the rate of increase of cylinder length $\ell$ is given by:

\begin{equation}
\frac{d\ell}{dt} = \alpha \left( \frac{4R}{3} + \ell \right),
\end{equation}

\noindent which results in the following growth equation for $\ell$:

\begin{equation}
\ell = e^{\alpha t} \left( \ell_0 + \frac{4R}{3} \right) - \frac{4R}{3},
\end{equation}

\noindent where $\ell_0$ is the initial cell cylinder length.

We treat cell division as an instantaneous conversion of the mother cell into two daughter cells that occupy the same total cell length. Specifically, the mother cell at position $\boldsymbol{r}$ with orientation $\boldsymbol{n}$ is replaced by two daughter cells of cylinder length $\ell_0$ at positions $\boldsymbol{r} \pm (R+\ell_{0}/2)\boldsymbol{n}$ with the same orientations $\boldsymbol{n}$. The condition that the daughter cells occupy the same total cell length as the mother cell requires division to occur at a final cell cylinder length $\ell = 2\ell_0 + 2R$. This treatment results in a doubling time $t_{\mathrm{double}}$:

\begin{equation} \label{eq:td}
t_{\mathrm{double}} = \frac{1}{\alpha} \log \left( \frac{10R + 6\ell_0}{4R + 3\ell_0} \right).
\end{equation}

\noindent Replacing the mother cell with two daughter cells in this manner results in a modest loss of total cell volume, but importantly, it does not increase the overlap with any neighboring cells. This division protocol was chosen to avoid introducing non-physical impulses that might alter reorientation dynamics.

\subsection*{Cell-to-cell repulsion}

Bacterial cells maintain their shape due to the presence of the cell wall. Although the cell wall itself is rigid, it is coated by soft materials such as cell-bound extracellular polysaccharides (EPS)\cite{S3}. These extracellular bio-components can deform elastically when cells encounter obstacles such as other cells or external surfaces, and these deformations produce repulsive pushing forces. To treat this elastic interaction, we employ the Hertzian theory of mechanical contact\cite{S1}. The elastic interaction between two cells, $i$ and $j$, has an energy that scales with the cell-cell overlap $\delta_{ij}$, defined for our model cells as $2R$ minus the smallest distance between the centerlines of the cell cylinders. For generic contact geometries of two spherocylinders, the contact energy is given by:

\begin{equation}
E_{\mathrm{cell-cell},ij} =E_0 R^{1/2} \delta_{ij}^{5/2},
\end{equation}

\noindent for $\delta_{ij}>0$ and $0$ for $\delta_{ij}<0$ (i.e. $0$ for cells not in contact), where $E_0$ is the cell stiffness.

\subsection*{Cell-to-surface interactions}

During biofilm growth, cells may interact with the surface. When a cell presses against the surface, the surface exerts a repulsive force against the cell. On the other hand, cells can secrete surface adhesion proteins Bap1/RbmC that coat the surface\cite{S3} and produce attractive forces. We therefore model cell-to-surface contact as a combination of repulsive and attractive interactions. To match the experimental surface geometry, which consists of a relatively flat and homogeneous surface, we model the surface as an infinite, two-dimensional plane located at $z=0$. We take the normal vector of the surface to point along the $z$-direction, which defines the vertical direction (Supplementary Fig. 1).

We treat the pushing interaction between cells and the surface analogously to the cell-to-cell interactions described above, but with a contact interaction that acts along the entire length of the cell. Specifically, the elastic contribution to the cell-to-surface contact energy is given by the integral of the elastic contact energy density along the centerline of the cell cylinder. In what follows, we parameterize the centerline as the set of points given by $\boldsymbol{r} + s \boldsymbol{n}$, where $r$ is the position of the cell center, $\boldsymbol{n}$ is a unit vector that specifies the cell orientation, and $s$ is a coordinate that runs from $-\ell/2$ to $\ell/2$ (Supplementary Fig. 1). Thus, the overlap $\delta(s)$ of each infinitesimal segment at $s$ with the surface is given by:

\begin{equation}
\delta(s)=R- (z+s \boldsymbol{n} \cdot \boldsymbol{\hat{z}}),
\end{equation}

\noindent for $\delta(s) >0$ and 0 otherwise (i.e. $0$ for points not in contact), where $z$ is the height of the cell center.

Furthermore, we also account for changes in the cell-to-surface contact geometry as the cell is reoriented (Supplementary Fig. 1). In the limit that the cells are completely horizontal, i.e., when $\boldsymbol{n} \cdot \boldsymbol{\hat{z}}=0$, we treat the contact geometry of the integrated surface interaction as that of the contact between a horizontal cylinder and a plane. For completely vertical cells, we treat the contact geometry of the integrated surface interaction as the contact between a sphere and a plane. For generic values of the cell orientation, the contact geometry is given by a sum of both cylindrical and spherical contributions weighted by a smooth crossover function that depends on the cell orientation $\boldsymbol{n}$, or equivalently, on the angle $\theta = \sin^{-1}(\boldsymbol{n} \cdot \boldsymbol{\hat{z}})$ between the cell and the surface. The crossover functions are chosen to be sinusoidal in $\theta$, as these are the simplest functions that preserve the scaling of contact energies with contact penetration for linear deviations around the horizontal and vertical orientations of the cell. Taken together, the contribution to the energy of cell $i$ due to its elastic cell-to-surface interactions is given by:

\begin{equation}
E_{\mathrm{el},i} = E_0 R^{1/2} \delta_{i}^{5/2},
\end{equation}

\noindent where $\delta_{i}^{5/2}$ is given by:

\begin{equation}
\delta_i^{5/2} = \int^{s/2}_{-s/2}  \left[  R^{-1/2} \cos^2 (\theta) \delta^2(s) + \frac{4}{3} \sin^2 (\theta) \delta^{3/2}(s) \right] ds.
\end{equation}

To model the cell-to-surface adhesion interaction\cite{S3}, we assume that each infinitesimal segment in contact with the surface provides a constant energy $-\Sigma_0$ per unit of contact area, according to the Derjaguin approximation\cite{S2}. The total contribution to the energy of cell $i$ due to cell-to-surface adhesion is given by:

\begin{equation} 
E_{\mathrm{ad},i} = -\Sigma_0 A_i,
\end{equation}

\noindent where $A_i = \int^{s/2}_{-s/2} a(s) ds$ is the total contact area as a function of the contact area density $a(s)$ given by:

\begin{equation}
a(s) = R^{1/2} \cos^2 (\theta) \delta^{1/2}(s) + \pi R \sin^2 (\theta) \Theta( \delta(s)) ,
\end{equation}

\noindent where $\Theta$ is the Heaviside step function. In the above expressions for $\delta_i^{5/2}$ and $a(s)$, we have incorporated the appropriate geometrical factors and scaling exponents for spherical and cylindrical Hertzian contacts. Thus, the total cell-to-surface energy $E_{{s},i}$ is given by:

\begin{equation} \label{eq:s}
E_{{s},i} = E_{\mathrm{el},i} + E_{\mathrm{ad},i}
\end{equation}

When all points of the cell's centerline are separated from the surface by distances larger than $R$, i.e. when the cell is detached from the surface, $E_{{s},i}$ is zero and the surface does not exert any force on the cell. In contrast, when the cell is in contact with the surface, the surface exerts both repulsive and attractive forces on the cell. In the absence of external forces, the competition between these opposing forces results in a stable fixed point at $\theta_0=0$, i.e., the cell is horizontal, and the penetration $\delta_0$ is given by:

\begin{equation}
\delta_0 = \frac{1}{R}\left( \frac{R^2 \Sigma_0}{4 E_0} \right)^{2/3}.
\end{equation}

\subsection*{Viscosity}

Cell motion is strongly opposed by drag from both its three-dimensional environment, including the surrounding ambient fluid and the polymer matrix, as well as by friction from the surface. For simplicity, we treat both of these effects via Stoke's drag terms that oppose the motion of each infinitesimal segment of the cell cylinder's centerline. For the ambient fluid, the density of the drag force along the centerline is taken to be proportional to $\eta_0 \boldsymbol{v}(s)$, where $\eta_0$ is the ambient viscosity and $\boldsymbol{v}(s)$ is the velocity of the segment at centerline position $s$:

\begin{equation}
\boldsymbol{v}(s) = \dot{\boldsymbol{r}} + s \dot{  \boldsymbol{\hat{n}}  },
\end{equation}

\noindent where the dot indicates the time derivative. To model the effect of the surface drag, we take the drag force provided by the surface to oppose the segment's motion tangential to the surface. Furthermore, we assume that the surface viscosity of a contacting segment is proportional to its contact area density $a(s)$ as given by the Hertzian contact geometry detailed above. The combination of ambient drag and surface drag corresponds to the following dissipation function:

\begin{equation} \label{eq:sd}
P_i = \frac{1}{2} \int_{-\ell /2}^{\ell / 2} \left( \eta_0 \boldsymbol{v}^2(s) + \frac{\eta_1 a(s)}{R} \left[ \boldsymbol{v}(s) - (\boldsymbol{v}(s) \cdot \hat{z}) \hat{z} \right]^2 \right) ds,
\end{equation}

\noindent where $\eta_1$ is the surface drag coefficient.

\subsection*{Equations of motion}

Taken together, the above interactions determine the equations of motion for the model cells. For a collection of cells, we define the total energy $E$ as follows:

\begin{equation}
E( \{ \boldsymbol{q}_i \} ) = \sum_i \left( E_{\mathrm{el},i} + E_{\mathrm{ad},i} \right)  + \sum_{ i \neq j } \left( E_{\mathrm{cell-cell},ij} \right),
\end{equation}

\noindent where $\boldsymbol{q}_i = \{\boldsymbol{r}_i, \boldsymbol{\hat{n}}_i \}$ is the generalized coordinate vector of cell $i$. We compute the equations of motion for cell $i$ using Lagrangian mechanics as follows:

\begin{equation}
\frac{\delta P_i}{\delta \dot{\boldsymbol{q}_i}} = -\frac{\delta E_i}{\delta \boldsymbol{q}_i} + \lambda \boldsymbol{\hat{n}}_i \frac{d\boldsymbol{\hat{n}}_i}{d\boldsymbol{q}_i},
\end{equation}

\noindent where $\lambda$ is a Lagrange multiplier introduced to account for the constraint $\boldsymbol{\hat{n}} \cdot \boldsymbol{\hat{n}} = 1$ on the cell orientation vector.

\subsection*{Choice of parameters}

We determined the parameters in our agent-based model by fitting them to experimental data:

\begin{itemize}
\item {Initial cell cylinder length $\ell_0$: In our cell model, the region enclosed by the spherocylinder represents the portion of the cell enclosed by the cell membrane and cell wall as well as the biopolymer coating that surrounds the cell. Since our experiment does not image the cell wall or coating directly, we determine the cell shape parameters $\ell_0$ and $R$ by fitting to simulations of the agent-based model. To do so, we identify the length of the rigid cell cylinder of the model cell with the length of the cell cylinder of the model cell plus an offset due to the presence of the biopolymer coating. Physically reasonable values of the offset lie between $0$ and $R$. Therefore, we report the cell cylinder length as the value in the center of this range, with the full range giving the error bars. In practice, we measure the initial cell cylinder length by first recording the average cell length from the experimental images, e.g. top row of Fig. 4a. We then determine the average cell length of the modeled cells by computing the average cell length as a function of the initial cell length $\ell_0$ (Fig. 2c). This function provides a mapping from the average experimental cell length to the initial cell length of the modeled cells.}
\item{ Cell spherocylinder radius $R$: We determine the radius $R$ from the experimental cell density near the edge of the biofilm (where cells are close-packed but under negligible compression). The cell radial density near the edge of the experimental biofilm in Fig. 1 is roughly $0.16$ cells per square micron (Fig. 3b), which is achieved by an agent-based model with $\ell_0=1.25R$ and $R=\SI{0.8}{\micrometer}$. The cell radius does not change significantly for the different drug conditions.}
\item{ Cell stiffness $E_0$: The cell stiffness is approximated as $E_0 = Y / (2-2\nu^2)$, where $Y$ is the Young's modulus, and $\nu$ is the Poisson ratio, in accordance with contact mechanics. These elastic parameters correspond to the effective material properties of the cell, which is a composite of the hard core of the cell starting at the cell wall and the soft biopolymer coating surrounding the cell. Since the cell wall is very rigid compared to the biopolymer coating, the elastic properties of cell interactions are primarily determined by the latter. The Young's modulus and Poisson ratio were measured using bulk rheology to be $Y\simeq \SI{450}{\pascal}$ and $\nu \simeq 0.49$.}
\item{ Cell growth rate $\alpha$: To model the noise in cell growth rate, we assign a random value of $\alpha$ to each cell upon birth. Specifically, we take $\alpha$ to be a random variable drawn from a Gaussian distribution. The mean value of $\alpha$ is determined from experiment by first measuring the average doubling time. The average doubling time for all the experimental colonies, (including those treated with drugs) is roughly $t_{\mathrm{double}} \sim 35$ minutes. From this value, we determine $\langle \alpha \rangle$ using Supplementary Eq. \ref{eq:td} above. The standard deviation of $\alpha$ is chosen to be $0.2 \langle \alpha \rangle$, to ensure that cell division events throughout the biofilm become desynchronized over times comparable to those observed in experiment.}
\item { Cell ambient viscosity $\eta_0$: We set the ambient viscosity $\eta_0$ equal to the biofilm viscosity measured from bulk rheology, which yields $\eta_0 \simeq \SI{20}{\pascal\second}$.}
\item { Cell surface drag coefficient $\eta_1$: We estimated $\eta_1$ using microfluidics. To do so, we inoculated cells in a microfluidic chamber at a low density, which allowed us to image isolated, individual cells adhered to the surface using high resolution microscopy. We subsequently gradually increased the flow rate until we observed cell motion. From the observed motion, we estimated the surface drag as $\eta_1 \sim F_{{s}} \Delta t / \Delta x$, where $F_{{s}}$ is the estimated force on the cell due to the effect of shear flow, $\Delta t$ is the duration of a short observation window, and $\Delta x$ is the distance traveled by the cell during the window. We estimate the force as $F_{{s}} = \eta_0 A (du/dz)$, $A$ is the cell footprint assuming a horizontal configuration and $du/dz$ is the derivative of the flow velocity in the direction along the channel, with the derivative calculated along the direction transverse to the surface and evaluated at the cell midline. This estimate yields a value $\eta_1 \simeq 2\cdot 10^5\ \SI{}{\pascal \second}$. }
\item { Cell adhesion $\Sigma_0$: The contact adhesion energy is chosen to match the onset time of verticalization for the modeled biofilm to that of the experimental biofilm (Fig. 1c,d). The corresponding value of $\Sigma_0$ yields a penetration depth $\delta_0 \simeq 0.04 R$. }
\end{itemize}

\newpage

\section{\textbf{Validation of quasi-3D approximation of the agent-based model}}

\begin{figure}[H]
\centering
\includegraphics[width=0.8\columnwidth]{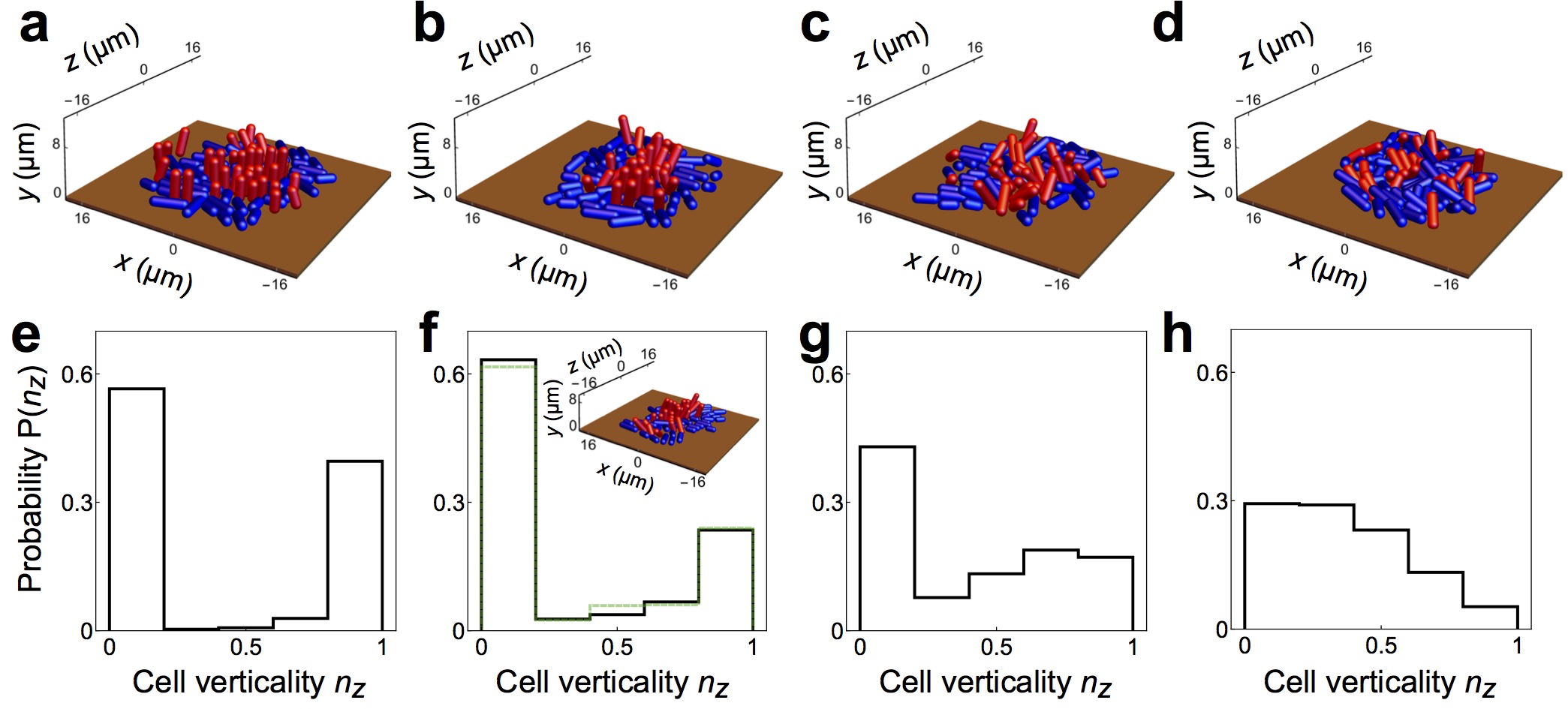}
\caption{\label{fig:FG2}
Varying the ratio of ambient viscosity to surface viscosity for modeled biofilms. (\textbf{a-d}) visualizations of full 3D agent-based model biofilms at time $t=250$ minutes, showing horizontal (blue) and vertical (red) cells and the surface (brown) for ambient viscosity $\eta_0 = \SI{20}{\pascal\second}$ and ratios $\eta_0 / \eta_1$ of ambient viscosity to surface viscosity (\textbf{a}) $10^{-5}$, (\textbf{b}) $10^{-4}$, (\textbf{c}) $10^{-3}$, and (\textbf{d}) $10^{-2}$. (\textbf{e-h}) Distributions $P(n_z)$ of cell verticality $n_z$ at time $t=250$ minutes for surface-adhered cells in the full 3D agent-based model biofilms in (\textbf{a-d}, black). The green curve in (\textbf{f}) shows $P(n_z)$ for the quasi-3D agent-based model biofilm reported in the main text. The inset of (\textbf{f}) shows visualization of the quasi-3D agent-based model biofilm at time $t=250$ minutes, showing horizontal (blue) and vertical (red) cells and the surface (brown) for ambient viscosity $\eta_0 = \SI{20}{\pascal\second}$ and ratio of ambient viscosity to surface viscosity $\eta_0/\eta_1 = 10^{-2}$.}
\end{figure}

Simulating mature three-dimensional biofilms requires prohibitively large amounts of computational time for systematic studies, due to the exponential growth of cells combined with the large separation of scales between the ambient and surface drag. However, for the purposes of describing the verticalization transition, it is sufficient to consider only the dynamics of the surface layer (see Results). Therefore, we developed a quasi-3D simplification of the full 3D agent-based model to make the computations tractable. This quasi-3D model exploits the large separation of scales between the ambient and surface drag by removing from the simulation cells that become detached from the surface. Since the cells on the surface are all subject to the large surface drag, the overall variation in forces throughout the biofilm is substantially reduced, which significantly lowers the computational time required to grow a biofilm surface layer of a given size. Provided the number of layers of cells above the surface layer is small, the forces exerted by these cells on the surface cells are negligible compared to the forces on surface cells produced by other surface cells, and so we expect this approximation to be accurate at early times.

To verify that this quasi-3D model is a reasonable simplification of the full 3D model, we directly compared the orientation patterning of both models for small biofilm sizes (Supplementary Fig. 2). We found that removing the detached cells results in a slightly narrower peak of the vertical cell orientation distribution. A simple explanation for this effect is that modest deviations of the cell orientation from a completely vertical orientation are not strongly constrained by the surface pressure, and thus the small forces exerted by detached cells in the full 3D simulations are enough to cause such deviations. To account for this feature, in simulations in which we remove surface-detached cells, we employ a larger effective value of the ambient viscosity $\eta_0 = 10^{-2} \eta_1$ to match the orientational distribution observed in the full, 3D simulations (Supplementary Fig. 2). This variant model can reproduce the orientational patterning of the surface layer in the full 3D model for smaller biofilms, as well as the orientational patterning of the surface layer in the experimental biofilm throughout the full duration of the experiment.

\newpage

\section*{\textbf{Supplementary Figures 3 and 4: Models for cell instabilities}}

Cell verticalization events are triggered by mechanical instabilities. In this section, we elucidate the physical mechanisms underpinning cell instabilities by investigating a series of minimal models. We first consider a line of cells under compression and show that instabilities are localized. Next, we study two different classes of dynamical instabilities that can occur at the cell-scale: surface compression and peeling. For each of these classes, we present a simplified rod-spring model followed by a more detailed model that includes the cell and surface geometries. Our detailed models describe the reorientation thresholds of the agent-based model biofilms (Fig. 2a,b).

\subsection*{Cell line instability}

Cells become unstable to verticalization under large enough compressive forces. This threshold effect is analogous to Euler buckling. In contrast to Euler buckling, however, we observed that cell instabilities are spatially localized within a biofilm cluster. A key feature underlying this discrepancy is the role played by the restoring potential, defined as the interaction that stabilizes the system in the absence of external forces. In Euler buckling, the restoring potential is provided by the rod's internal bending rigidity, whereas for cell instabilities, the restoring potential is provided by the external surface. Therefore, to understand why verticalization is localized, we start by briefly reviewing the conventional scenario of Euler buckling before going on to study the verticalization instability of a biofilm.\newline

\textbf{Case I: Euler buckling}\newline

The energy of an inextensible elastic rod under uniform compression is approximated by\cite{S1}:

\begin{equation}
E_{\mathrm{rod}} = \frac{1}{2} \int dx \left[ \kappa \left( \frac{d^2h(x)}{dx^2} \right)^2 - F \left( \frac{dh(x)}{dx} \right)^2   \right],
\end{equation}

\noindent where $h(x)$ is the height field (transverse to the rod's axis), $\kappa$ is the bending rigidity, and $F$ is the externally applied compressive load. Here, the first term is the restoring potential and the second term corresponds to the work performed by the external force, which is proportional to the end-to-end contraction $\Delta x \simeq \int (dh/dx)^2$. The Fourier decomposition of the height field is given by:

\begin{equation}
h(x) = \sum_q h_q \sin(q x),
\end{equation}

\noindent where $h_q$ is the amplitude of a mode with frequency $q$. Modes with $F>\kappa q^2$ provide a negative contribution to the energy and thus are unstable. If the force $F$ is increased from zero, the first mode to become unstable is the lowest spatial frequency mode. In this case, the lowest mode $q_1$ is an extended deformation limited by the length $\ell_{\mathrm{tot}}$ and it corresponds to $q_1 = \pi / \ell_{\mathrm{tot}}$ (Supplementary Fig. 3a).\newline

\begin{figure}[H]
\centering
\includegraphics[width=0.8\columnwidth]{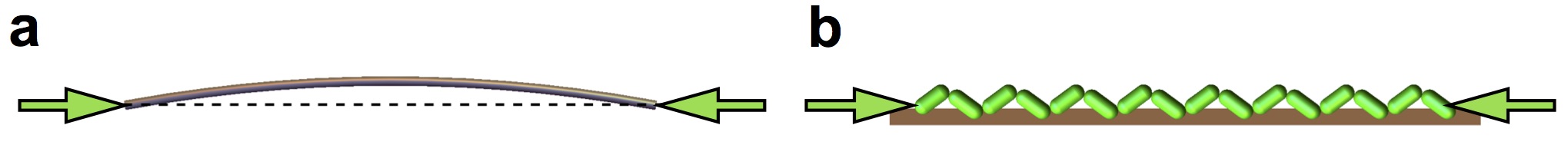}
\caption{\label{fig:FG3}
Compressive instabilities of one-dimensional media. (\textbf{a}) Schematic of Euler buckling of an elastic rod under uniform compression. As the equal-and-opposite compressive forces are increased from zero, the rod first becomes unstable to an extended deformation given by the lowest spatial frequency mode. (\textbf{b}) Schematic of the verticalization instability of a line of cells under uniform compression. As the compressive forces are increased from zero, the line of cells first becomes unstable to a combination of vertical motions and rotations, with a large mode number.
}
\end{figure}

\textbf{Case II: Cell line verticalization}\newline

We now consider a cluster of cells interacting with a surface. For brevity, we focus on a line of cells in one dimension. Thus, each cell $i$ can undergo center of mass motion transverse to the line, as well as rotate. For stiff cells ($E_0 \rightarrow \infty$), the cell-cell contact distance remains fixed. In the continuum limit of small cells, the end-to-end contraction $\Delta x$ of the biofilm is given by:

\begin{equation}
\Delta x \simeq \frac{1}{2} \int dx \left[ c_1 \left(  \frac{dh(x)}{dx}\right)^2 + c_2 \theta(x)^2 - c_3 \theta(x) \frac{dh(x)}{dx}  \right],
\end{equation}

\noindent to second order in the height field $h(x)$ and the orientation field $\theta(x)$, where $c_1$, $c_2$, and $c_3$ are geometrical factors on the order of the cell length. Intuitively, these terms arise because both differential changes in cell heights as well as cell rotations (first and second terms) free up space along the surface and allow the cluster to pack more densely. However, coupled center of mass motions and rotations can either increase or decrease the contraction depending on their signs (third term).

The continuum limit of the surface energy (Supplementary Eq. \ref{eq:s}) is given by:

\begin{equation}
E_{\mathrm{ad}} = \frac{1}{2} \int dx \left[ \lambda_1 h^2(x) + \lambda_2 \theta^2(x)   \right],
\end{equation}

\noindent where $\lambda_1$ and $\lambda_2$ are elastic parameters proportional to the cell stiffness $E_0$ in the limit of small penetration $E_0 \gg \Sigma_0 R^{-1}$. Thus, for a biofilm under a uniform compressive load $F$, the total energy is given by:

\begin{equation}
E_{\mathrm{col}} = \frac{1}{2} \int dx \left[ \lambda_1 h^2(x) + \lambda_2 \theta^2(x)  - F c_1 \left(  \frac{dh(x)}{dx}\right)^2 + F c_2 \theta(x) \frac{dh(x)}{dx} - F c_3 \theta(x)^2   \right].
\end{equation}

\noindent To understand how this line of cells becomes unstable, it is instructive to first consider what happens when either rotations or center of mass motions are forbidden. The former scenario corresponds to a flexible chain. Here, modes with $F>\lambda_1 / (c_1 q^2)$ are unstable, so the instability first occurs through the \emph{highest} spatial frequency mode. On the other hand, when center of mass motions are forbidden, the biofilm first becomes unstable when $F> \lambda_2 / c_2$, independent of the mode number.

When both center of mass motions and rotations are allowed, the coupling between the height field and the orientation field can facilitate the instability. In particular, when both fields are completely out of phase, the negative contribution to the energy due to the coupling is maximized. As the force is increased from zero, the first unstable mode is therefore a combination of center of mass motions and rotations at a large mode number (Supplementary Fig. 3b). Thus, in contrast to Euler buckling, cells first become unstable to verticalization on length scales comparable to the cell length.

How do these results apply to growing biofilms? In a growing biofilm, cells are subject to a spatially non-uniform distribution of forces. Since verticalization instabilities can proceed on wavelengths comparable to cell length, any region in which growth-derived forces overcome the restoring potential will become locally unstable. Thus, the propensity for verticalization to occur at high mode number explains why we observed cell reorientations to occur locally in regions of large forces (Fig. 2d).

\subsection*{Toy model for compression instability}

Our results for the line of cells implies that verticalization instabilities occur at the single-cell scale. Therefore, to understand the onset of reorientation, we now turn to models for the instabilities of individual cells under applied forces. We first explore a minimal toy model that consists of a rigid rod of length $\ell$ attached to an elastic foundation. The elastic foundation is comprised of a large number of identical Hookean springs spread evenly over the length of the rod (Supplementary Fig. 4a). The ends of the springs are fixed to lie at the same height, which allows for an unstretched reference configuration with elevation angle $\theta=0$. We consider motions for which the rod is free to rotate about its center, i.e. to finite values of $\theta$. Thus, the energy $E_{\mathrm{ef}}$ of the elastic foundation is given by:

\begin{equation}
E_{\mathrm{ef}} = \frac{k \ell^3}{2} \sin^2 \theta,
\end{equation}

\noindent to leading order in $\theta$ in the limit of a continuous foundation, where $k$ is an elastic parameter with the same units as the cell stiffness $E_0$. To represent the cell-cell interactions, we apply equal-and-opposite forces of magnitude $F$ to both ends of the rod. The forces act to squeeze the rod and always point along the initial direction of the rod. Thus, the total energy of the system $E_{\mathrm{tot}}$ is given by:

\begin{equation}
E_{\mathrm{tot}} = \frac{k \ell^3}{2} \sin^2 \theta - F \ell \cos \theta,
\end{equation}

\noindent For $F=0$, the rod rests on the foundation at an elevation angle $\theta=0$. For motion around this configuration opposed by friction, the elastic and compressive forces must balance the drag force $F_{d}$, which is proportional to the rate of change of the elevation angle:

\begin{equation}
F_d \sim \dot{\theta},
\end{equation}

\noindent assuming that the friction is provided by Stoke's drag terms that act along the length of the rod, as in Supplementary Eq. \ref{eq:sd}. Thus, for small elevation angles, the rate of change of $\theta$ is given by:

\begin{equation}
\dot{\theta} \sim (F - k \ell^2) \theta.
\end{equation}

\noindent For small forces, $\theta=0$ is a stable fixed point of the system. However, when the force $F$ becomes large, the rod becomes unstable to reorientation. This bifurcation instability occurs at a threshold force $F_t = k\ell^2$. Intuitively, the dependence of this threshold force on cell length arises because the elastic foundation provides a fixed restoring energy per unit length of the rod.

\begin{figure}[H]
\centering
\includegraphics[width=0.8\columnwidth]{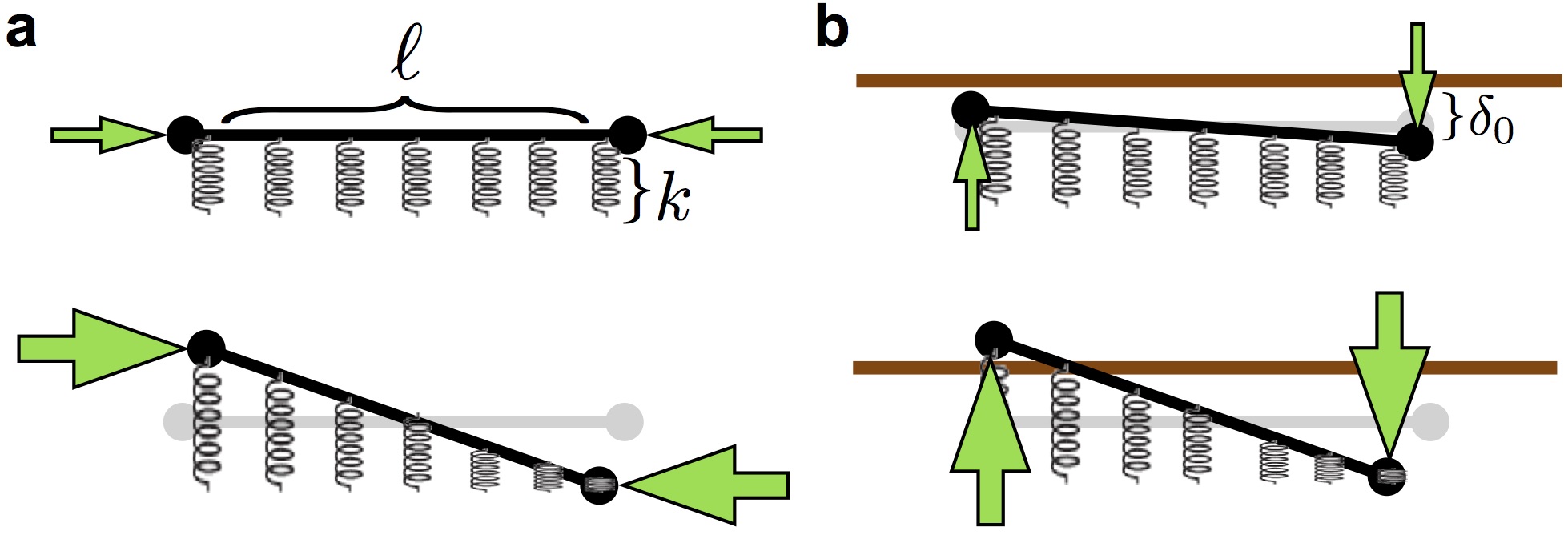}
\caption{\label{fig:FG4}
Toy models for cell-scale mechanical instabilities. (\textbf{a}) Schematic depiction of the toy model for surface compression instability, showing a rigid rod of length $\ell$ on an elastic foundation with a vertical stiffness modulus $k$. The rod is compressed by external, equal-and-opposite forces $F$ oriented horizontally (green arrows). For small values of force $F<F_{t} \sim k\ell^2$ (top), the rod remains stationary. For large values of force $F>F_{t}$ (bottom), the rod becomes unstable to reorientation. (\textbf{b}) Schematic depiction of the toy model for peeling instability, showing the system from (\textbf{a}) with the rod initially embedded a distance $\delta_0$ below the surface (brown). The rod is under an external torque $\tau$ (green arrows). For small values of torque $\tau<\tau_{t} \sim k\ell^2$ (top), the equilibrium position of the rod shifts to finite elevation angles. For large values of torque $\tau>\tau_{t}$ (bottom), one end of the rod moves above the surface. If we assume that the elastic foundation detaches from the segment of the rod above the surface, the rod will be unstable to further reorientation.
}
\end{figure}

\subsection*{Instability of model cell to compression}

The squeezed rod model shows how horizontal forces can result in the vertical motion of a rigid object. However, we expect this instability mechanism to depend on the details of the cell geometry, as well as the cell-to-surface interactions. Therefore, we now perform an analogous stability analysis on our spherocylindrical model cells in the presence of the surface potential given by $E_{{s},i}$ in Supplementary Eq. \ref{eq:s}.

In what follows, we allow the cell to undergo vertical center of mass motion in addition to rotation. In the absence of forces, the cell is embedded in the surface at its stable fixed point ($\theta=0$ and $\delta = \delta_0$). To mimic the average distribution of forces acting on cells in a biofilm, we now consider applying a uniform distribution of forces around the cell perimeter and acting in the $xy$ plane. For simplicity, we take the force at each point along the perimeter to be applied by a spherical piston of radius $R$ (equal to the cell radius) that is rigid with respect to the cell. The centers of the pistons are fixed to lie at the same height $R-\delta_0$ as the midline of the cell in the initial configuration, and their motions in the $xy$ plane are constrained to occur entirely along the direction of the shortest line connecting their centers to the cell cylinder's centerline in the initial configuration. For concreteness, we take the initial cell orientation vector to point in the $\boldsymbol { \hat { x}} $ direction. For pistons applied to the endcaps of this cell, the vector $\boldsymbol{d}_{\mathrm{end},\pm}$ between the piston's center and the cell cylinder's centerline is given by:

\begin{equation}
\boldsymbol{d}_{\mathrm{end},\pm} =  ( R - \delta_0 - z ) \boldsymbol{\hat{z} } + (\ell/2) \boldsymbol{\hat{n}} \pm (2R - \Delta d_{\mathrm{end},\pm} ) \boldsymbol{\hat{m}} \pm (\ell/2) \boldsymbol{\hat{x}}  ,
\end{equation}

\noindent where $z$ is the cell height above the surface, $ \boldsymbol{\hat{n}} = (\cos \theta, 0, \sin \theta)$ is the cell orientation vector, $ \boldsymbol{\hat{m}} = (\sin \phi, 0, \cos \phi)$ is the piston's angle of attack, and $ \Delta \boldsymbol{d}_{\mathrm{end},\pm} $ is the piston's displacement. As the cell moves, the piston is assumed to stay in contact with the cell. This constraint corresponds to the following equation:

\begin{equation}
|\boldsymbol{d}_{\mathrm{end},\pm}| = 2R,
\end{equation}

\noindent which determines $ \Delta d_{\mathrm{end},\pm} $ as a function of the cell configuration. For a piston applied to the cylindrical portion of the cell, the vector $\boldsymbol{d}_{\mathrm{side},\pm}$ between the piston's center and the cell cylinder's centerline is given by:

\begin{equation}
d_{\mathrm{side},\pm} = ( R - \delta_0 - z ) \boldsymbol{\hat{z} } - (2R - \Delta d_{\mathrm{side},\pm} ) \boldsymbol{ \hat{ y }}  + s \boldsymbol{ \hat{ x }} .
\end{equation}

\noindent The constraint that the piston remains in contact with the cell is specified by the following equation:

\begin{equation}
|\boldsymbol{d}_{\mathrm{side},\pm} - (\boldsymbol{d}_{\mathrm{side},\pm} \cdot \boldsymbol{\hat{n}} ) \boldsymbol{\hat{n}} | = 2R,
\end{equation}

\noindent which determines the piston displacement $\Delta d_{\mathrm{side},\pm}$ as a function of the cell configuration. The total work $W_p$ performed by the pistons on the cell is obtained by integrating the contributions from pistons around the perimeter of the cell:

\begin{equation}
W_p = p \int_0^{\pi} d\phi ( \Delta d_{\mathrm{end},+} +   \Delta d_{\mathrm{end},-}  ) +  p \int_{-\ell/2}^{\ell/2} ds ( \Delta d_{\mathrm{side},+} +   \Delta d_{\mathrm{side},-}  ),
\end{equation}

\noindent where $p$ is the applied ``pressure''. The total energy $E_{\mathrm{cp}}$ of the cell-piston system, i.e. the cell-to-surface energy minus the work done by the pistons on the cell, is given by:

\begin{equation}
E_{\mathrm{cp}} = E_{{s,i}} - W_p.
\end{equation}

\noindent For motion around this configuration opposed by friction, the equations of motion are given by:

\begin{equation}
\dot{z} = -\frac{1}{\eta_0 \ell} \frac{\partial E_{\mathrm{cp}}}{\partial z},
\end{equation}

\begin{equation}
\dot{\theta} = -\frac{12}{\eta_0 \ell^3} \frac{\partial E_{\mathrm{cp}}}{\partial \theta},
\end{equation}

\noindent to leading order in $z$ and $\theta$, where we have assumed that the friction is determined according to the dissipation function Supplementary Eq. \ref{eq:sd}. To determine the behavior of the system as a function of the applied surface pressure, we perform a linear stability analysis around the initial configuration. We first construct the stiffness matrix $\mathcal{D}$ as follows:

\begin{equation}
\mathcal{D} = \begin{bmatrix}
     -\frac{1}{\eta_0 \ell} \frac{\partial^2E_{\mathrm{cp}}}{\partial z^2}       &  -\frac{1}{\eta_0 \ell} \frac{\partial^2E_{\mathrm{cp}}}{\partial z \partial \theta} \\
    -\frac{12}{\eta_{0} \ell^3} \frac{\partial^2E_{\mathrm{cp}}}{\partial \theta \partial z}       & -\frac{12}{\eta_{0} \ell^3} \frac{\partial^2E_{\mathrm{cp}}}{\partial \theta^2}
  \\
\end{bmatrix}.
\end{equation}

\noindent For our model cell under surface pressure from the pistons, the off-diagonal terms of this matrix are zero, which indicates that vertical motion is decoupled from rotation. Therefore, the signs of the diagonal terms determine whether the cell is stable to infinitesimal perturbations. For our choice of parameters above (Supplementary Fig. 1) and for small values of force, both eigenvalues are negative and the system is stable. However, as the force is increased, the cell first becomes unstable to reorientation. The threshold value of surface pressure $p_{t}$ for which this instability occurs is given by:

\begin{equation}
p_{t} = \frac{3 E_0 \ell^2 R - 8 (3 \pi - 1) \Sigma_0 + 9 \sqrt[3]{2} R^{-1} \Sigma_0^{4/3} E_0^{-1/3} }{\ell^2+3 \pi \ell R+24 R^2} .
\end{equation}

\noindent For physiologically-relevant parameters, this surface pressure is roughly linear as a function of $\ell$ over a large range around $\ell = R$. For $\ell \sim R$, the threshold surface pressure is approximately:

\begin{equation}
p_{t} \sim E_0 (b_1 \ell - b_2 R),
\end{equation}

\noindent in the limit of small penetration (cell stiffness $E_0 \gg \Sigma_0 R^{-1}$), where $b_1 = (144+9\pi)/(25+3\pi)^2$ and $b_2 = 69R/(25+3\pi)^2$. The dependence of the threshold surface pressure on cell length arises in this regime because the total forces acting on the cell endcaps are comparable to the total forces acting on the cell cylinder. For longer cell lengths, however, the forces acting on the cell cylinder dominate and the threshold surface pressure saturates to $p_{t} \sim 3 E_1 R$. Intuitively, this saturation occurs because the pistons provide a fixed surface energy per unit length that balance the fixed surface energy per unit length of the model cell.

In the intermediate cell length regime, the scaling $p_{t} \sim R$ for $\ell \sim R$ implies that $F_{t} \sim R^2$, as in the toy model. However, the spherocylindrical cell model deviates from the toy model in two compensating ways. First, the work performed by the pistons to rotate the spherocylindrical cell scales more rapidly with cell length than the work performed by the purely horizontal forces in the toy model. For the case of forces applied to the end of the cell, the piston yields $W_{p} \sim \ell^2 \theta^2$ whereas the horizontal forces yield $W_{p} \sim \ell \theta^2$. Second, for a fixed amount of total force, spreading the pistons around the entire perimeter of the cell yields a smaller amount of in-plane torque than if the forces were concentrated entirely at the ends, as in the toy model. Thus, our spherocylindrical cell model demonstrates that it is important to consider the full effects of the cell-cell contact geometry together with the cell-cell contact distribution to fully capture the surface compression instability.

\subsection*{Toy model for peeling instability}

Our agent-based simulations suggest that for long cell lengths, forces in the $z$ direction play an important role in triggering verticalization. To describe this effect, we now return to the toy model of a rod on an elastic foundation discussed above and we consider the effect of external forces in the $z$ direction (Supplementary Fig. 4b). For simplicity, we take the rod's center of mass to be fixed. In this case, the configuration of the rod depends on the net torque $\tau$ provided by the external forces. The total energy $E_{z} $ of the system becomes:

\begin{equation}
E_{z} = \frac{k \ell^3}{2} \sin^2 \theta - \tau \theta.
\end{equation}

\noindent Upon minimizing this energy, we find that the applied torque shifts the stable configuration of the cell to a finite elevation angle $\theta_0 = \tau / k \ell^3$. How would this finite elevation angle influence the contact between a cell and the surface? For large elevation angles, the bonds between a cell and the surface must eventually break. When this occurs, continued peeling of the cell from the surface requires decreasing amounts of external torque. We can incorporate this mechanism into the torqued rod model in a simple manner by assuming the springs of the elastic foundation break when they are stretched more than a small distance $\delta_{t}$. For $\delta_{t} \ll \ell$, this distance is reached by one end of the cell when $\theta_0 \simeq \delta_{t} / \ell$. Therefore, the threshold torque for peeling scales as $\tau_{t} \sim \ell^2$.

\subsection*{Instability of model cell to peeling}

To determine the threshold verticalization torque for the model cell, we consider the spherocylindrical model cell in the presence of the surface potential. For simplicity, we take the cell center to remain fixed. For a small constant torque $\tau$, the stable angle $\theta_0$ is obtained by solving the following equation:

\begin{equation}
\tau = \frac{\partial   E_{s} } {\partial \theta} \Bigm| _{\theta_0}.
\end{equation}

\noindent For $\theta_0 \ll 1$, we find that $\theta_0 = \tau / b_3$, where $b_3$ is given by:

\begin{equation}
b_3 = (3 E_0 \ell^3 - 8 (3 \pi - 1) \Sigma_0 R^{-1} \ell + 9 \sqrt[3]{2} R^{-2} \ell \Sigma_0^{4/3} E_0^{-1/3}) / 12 .
\end{equation}

\noindent The critical angle $\theta_c$ for peeling a cell from the surface is reached when one end of the cell begins to leave the surface, i.e.:

\begin{equation}
\delta_0 \simeq \frac{\ell}{2} \theta_{t}.
\end{equation}

\noindent Setting $\theta_0 = \theta_{t}$ yields a threshold torque $\tau_{t} = 2 b_3 \delta_0 / \ell$. In the limit of small penetration (cell stiffness $E_0 \gg \Sigma_0 R^{-1}$), $\tau_{t}$ is given by:

\begin{equation}
\tau_{t} = \delta_0 E_0 \ell^2 / 2 .
\end{equation}

\noindent Thus, in this regime we find that $\tau_{t} \sim \ell^2$, in agreement with the scaling found for the torqued rod.

\newpage

\section*{\textbf{Supplementary Figures 5 to 7: Verticalization events (in experiment and simulation)}}

\textbf{Tracking verticalization events} To probe the local conditions driving verticalization, we tracked the cell-to-cell contact forces acting on individual cells in the agent-based model around the time when cells start to become vertical (Supplementary Fig. 5). We found that as a cell becomes vertical, the total surface force acting on it reaches a local maximum before decaying rapidly. The trend arises due to the nonlinearity of the cell-to-surface contact geometry, combined with the reduced footprint taken up by a vertical cell relative to a horizontal cell. That is, before the cell starts to become vertical, the surface forces increase due to cell growth, which increases cell-cell overlaps more rapidly than the overlaps can be resolved by rearrangements of cells. As the cell becomes vertical, it requires progressively lower amounts of force to induce further reorientation due to the peeling of the cell (see \textbf{Peeling instability of model cell} above). Reorientation frees up space along the surface for local rearrangements that reduce cell-cell overlaps and thereby alleviate the accumulating forces. These complex dynamics are readily apparent from visualizations of the force chains throughout a biofilm (Supplementary Video 3).

As a result of this behavior, the forces acting on a cell provide a characteristic signature of its transition from horizontal to vertical. Specifically, we identified the moment $t_{{r}}$ of the verticalization transition as the time of the peak force prior to the cell exceeding a critical orientation, which we took to be $n_z > 0.25$. In the main text, we showed that the values of the peak forces are consistent with the instability models we presented above (Fig. 2, Supplementary Fig. 3,4). \\

\textbf{The effect of the cell-cell contact distribution on the predicted reorientation pressure} In the main text and in a section above, we presented a theoretical prediction for the average reorientation pressure $\langle p_r \rangle$ in the agent-based model obtained by performing a linear stability analysis for a modeled cell under uniform surface pressure (see \textbf{Instability of model cell to compression} above). We found a large discrepancy between the calculated threshold reorientation pressure $p_t$ and the observed average reorientation pressure $\langle p_r \rangle$ (Fig. 2a). To eliminate the possibility that the discrepancy could arise from heterogeneity in the contact forces in the agent-based model, we made a separate theoretical prediction for $\langle p_r \rangle$ that incorporates the numerically-observed distribution of cell-cell contact forces. Specifically, for each reorientation event, we first recorded the distribution of cell-cell contact forces, i.e. the magnitudes, directions, and points of application of forces in the $xy$ plane applied to the reorienting cell by neighboring cells. For each set of cell-cell contact forces, we determined the threshold surface pressure via linear stability analysis by uniformly rescaling the magnitudes of the forces until the onset of an instability. Incorporating the numerically-observed distribution of cell-cell contact forces in this manner did not yield a substantially different prediction for $\langle p_r \rangle$ compared to the prediction assuming a uniform surface pressure (Supplementary Fig. 6). Based on the substantial discrepancy between $p_t$ and $\langle p_r \rangle$, we hypothesized that cell-cell forces acting along the $xy$ plane alone do not account for the verticalization of long cells.

\newpage

\begin{figure}[H]
\centering
\includegraphics[width=0.8\columnwidth]{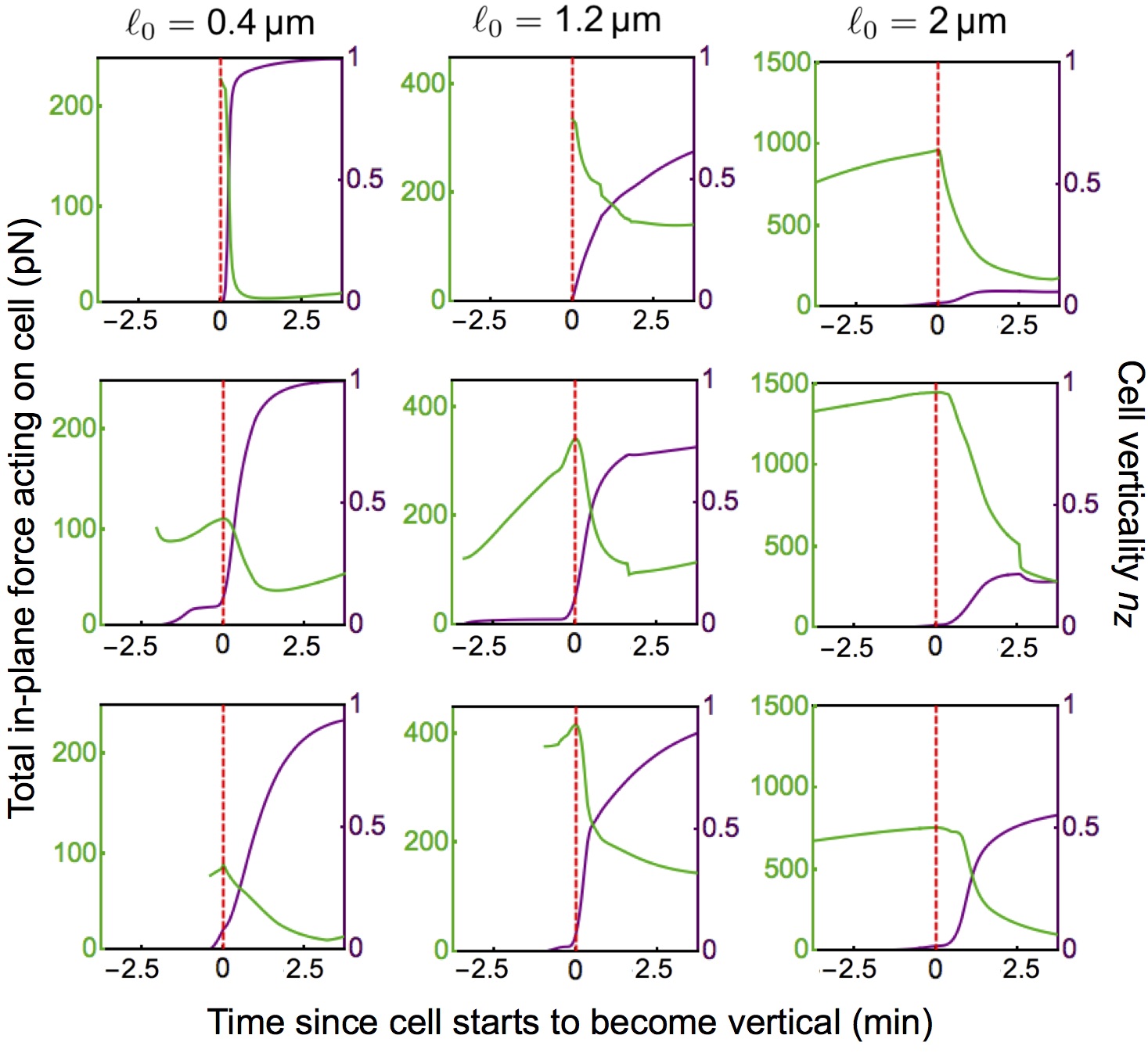}
\caption{\label{fig:FG9002}
Verticalization of individual cells in the agent-based model. Representative examples of the total surface force on a cell (green), defined as the total contact force in the $xy$ plane acting on a cell, and cell verticality $n_z$ (purple) versus the time since the cell starts to become vertical (red vertical dashed line), for cell cylinder lengths $\ell_0 = \SI{0.4}{\micrometer}$ (left column), $\ell_0 = \SI{1.2}{\micrometer}$ (middle column), and $\ell_0 = \SI{2}{\micrometer}$ (right column). For $\ell_0 = \SI{0.4}{\micrometer}$ and $\ell_0 = \SI{1.2}{\micrometer}$, the traces begin at the moment of cell birth, whereas only partial traces are shown for $\ell_0 = \SI{2}{\micrometer}$.
}
\end{figure}

\begin{figure}[H]
\centering
\includegraphics[width=0.4\columnwidth]{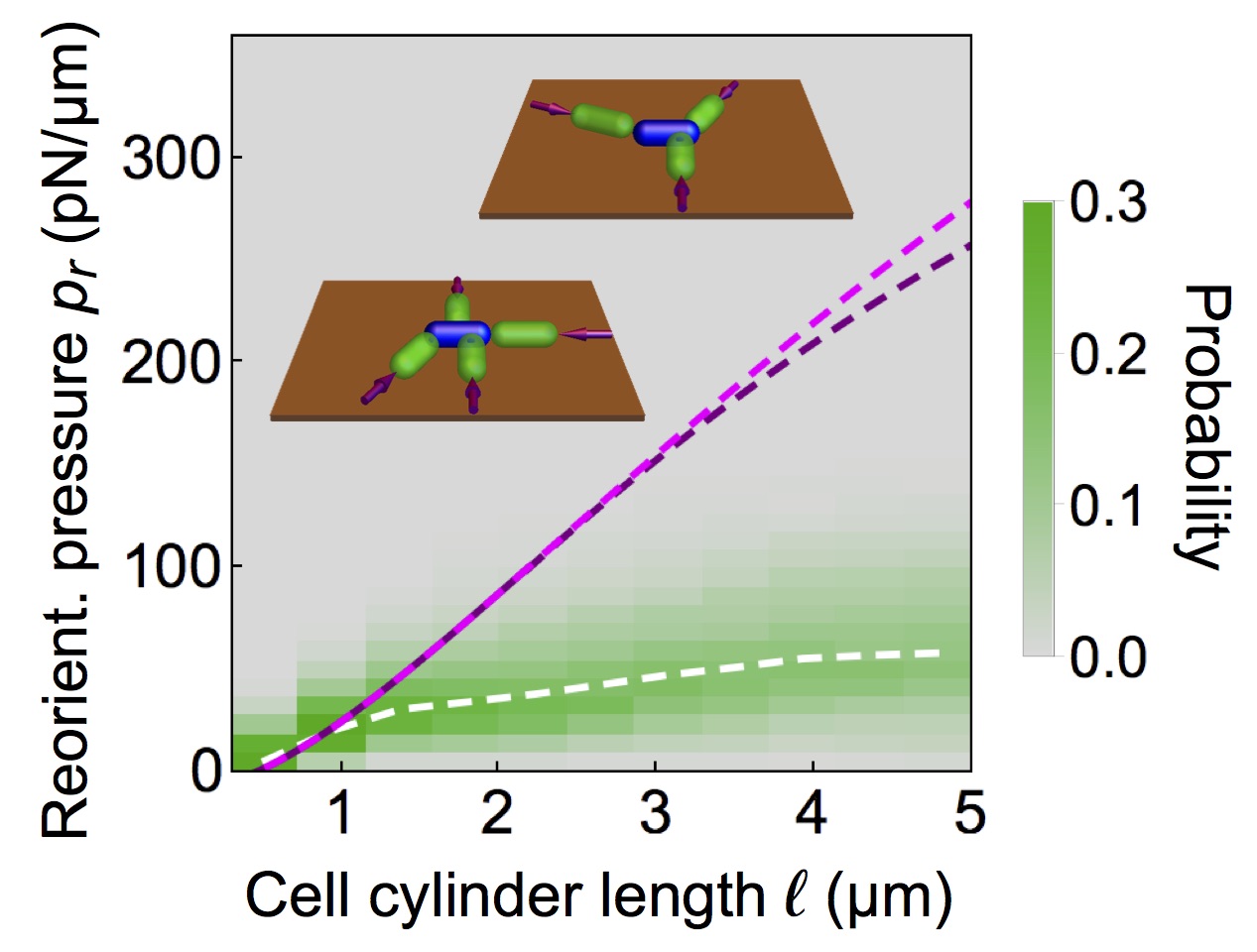}
\caption{\label{fig:FG9002}
Incorporating the distribution of cell-cell contact forces into the compression instability model. Distributions of reorientation surface pressure $p_{r}$, defined as the total contact force in the $xy$ plane acting on a cell at time $t_{r}$ of verticalization, normalized by the cell's perimeter, versus cell cylinder length $\ell$. White dashed curve shows the average reorientation surface pressure $\langle p_{r} \rangle$ as a function of $\ell$. Magenta dashed curve shows theoretical prediction for $\langle p_{r} \rangle$ from linear stability analysis for a modeled cell under uniform surface pressure, and purple dashed curve shows average of the predicted distribution of $p_{r}$ from linear stability analyses for a sample of reorienting modeled cells under the numerically-observed cell-cell contact forces. The numerical data for $p_r$ and the distribution of contact forces is obtained from all reorientation events among different biofilms simulated for a range of initial cell lengths $\ell_0$. Insets show schematic depictions of example cell-cell contact geometries considered in the linear stability analysis.
}
\end{figure}

\begin{figure}[H]
\centering
\includegraphics[width=0.8\columnwidth]{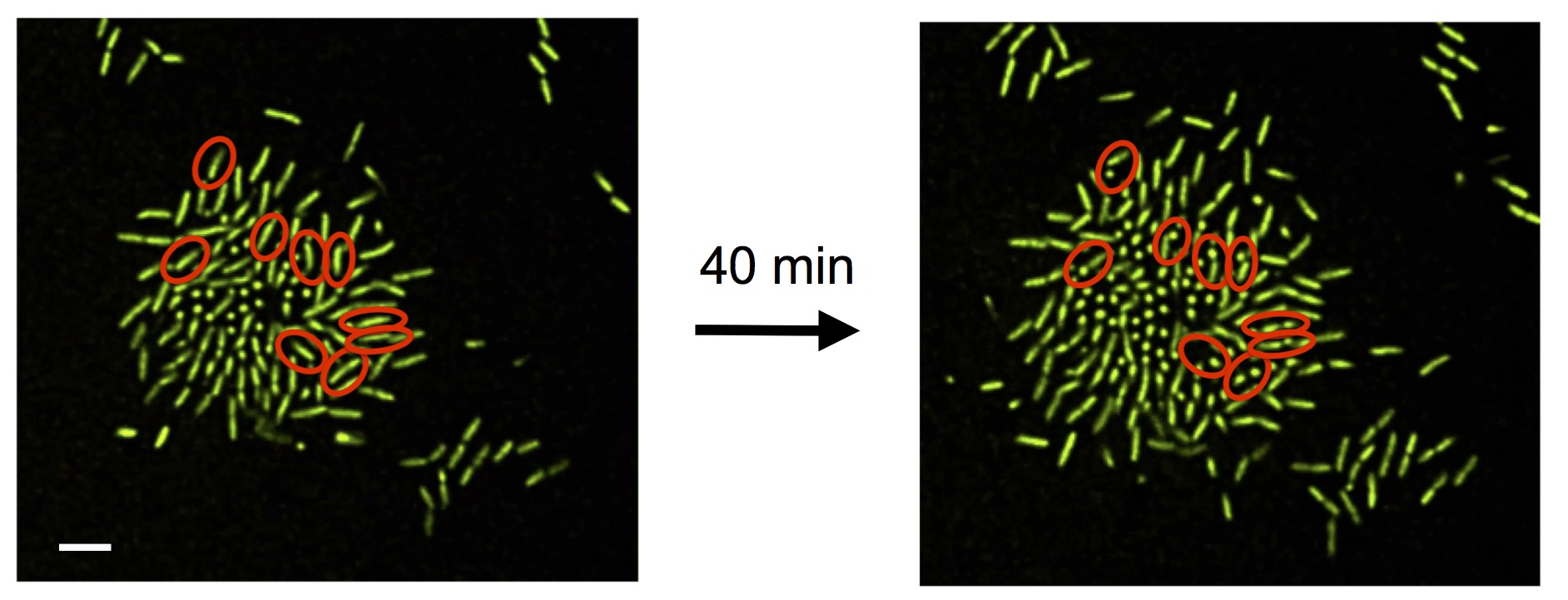}
\caption{\label{fig:FG9003}
Division-triggered reorientation events. Confocal fluorescence microscopy images of living, growing biofilm under standard conditions at approximately $t=300$ minutes (left) and $40$ minutes later (right). Red circles indicate mother cells immediately prior to division (left) with either one or both daughter cells becoming vertical following division (right). Scale bar: $\SI{5}{\micrometer}$.
}
\end{figure}

\newpage

\section*{\textbf{Supplementary Figures 8 to 10: Cell avalanches}}

How are cell verticalization events correlated in space and time? To quantify such correlations, we computed the joint radial distribution $P(\Delta r_{ij} , \Delta t_{{r},ij} ) / \Delta r_{ij}$. Here, $P(\Delta r_{ij} , \Delta t_{{r},ij} )$ is the joint distribution of spatial separations $\Delta r_{ij}$ and temporal separations $\Delta t_{{r},ij}$, where $\Delta r_{ij} = | \boldsymbol{r}_i - \boldsymbol{r}_j |$, where $\boldsymbol{r}_i$ is the position of cell $i$ at the time $t_{{r},i}$ of the peak of total force on the cell prior to it becoming vertical, and $\Delta t_{{r},ij} = |t_{{r},i} - t_{{r},j}|$ (Supplementary Fig. 8). For all values of average cell length we studied, this distribution of separations displayed a characteristic peak for small spatial and temporal separations, followed by a rapid decay in both distance and time.

\begin{figure}[H]
\centering
\includegraphics[width=0.8\columnwidth]{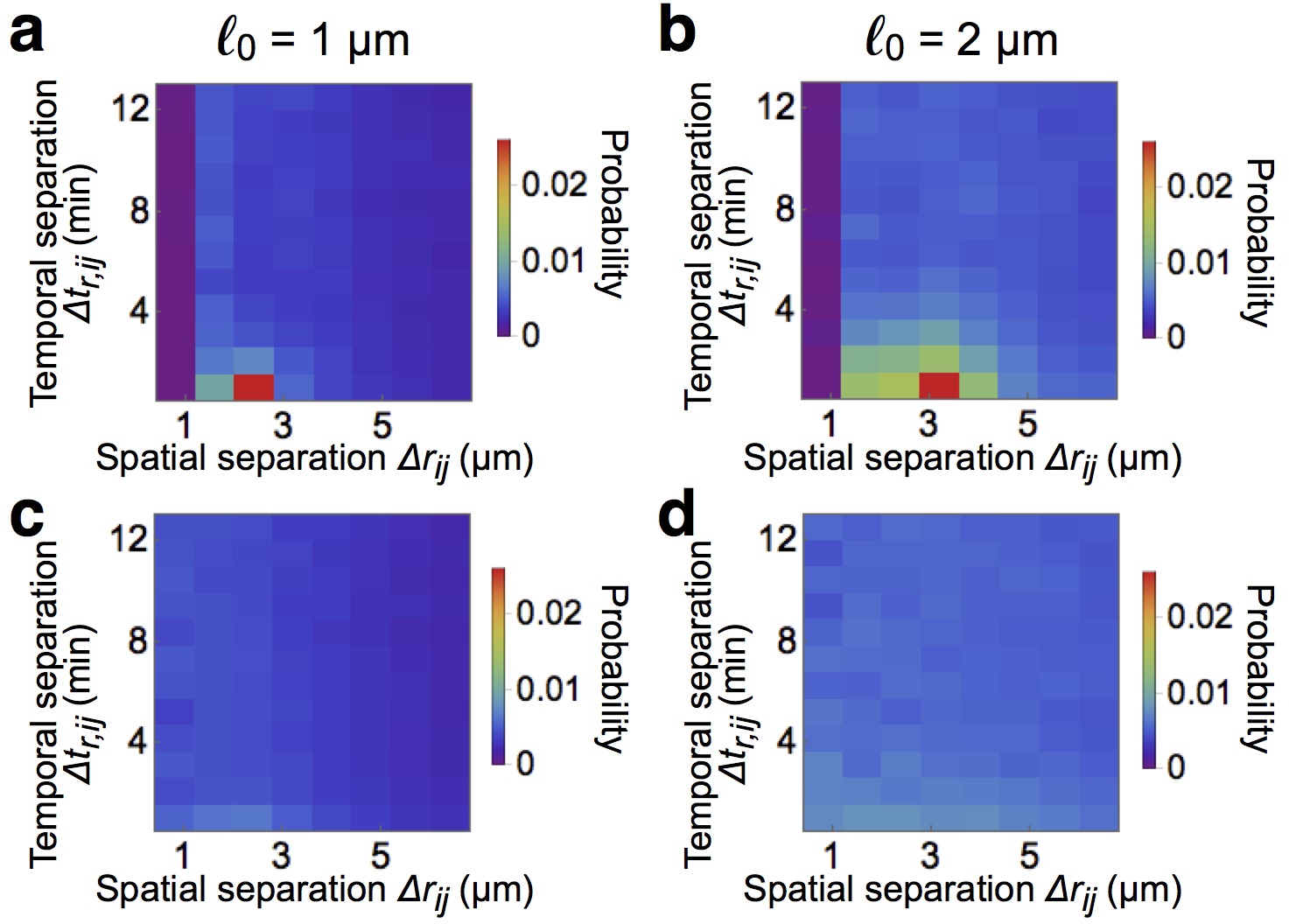}
\caption{\label{fig:FG6}
Correlations among verticalization events for agent-based model. (\textbf{a-b}) Joint radial distribution $P(\Delta r_{ij} , \Delta t_{r,ij} ) / \Delta r_{ij}$ of spatial separations $\Delta r_{ij} = | \boldsymbol{r}_i - \boldsymbol{r}_j |$, where $\boldsymbol{r}_i$ is the position of cell $i$ at the time $t_{r,i}$ of reorientation, and temporal separations $\Delta t_{ij} = |t_{r,i} - t_{r,j}|$ between pairs of reorientation events for (\textbf{a}) $\ell_0 = \SI{1}{\micrometer}$ and (\textbf{b}) $\ell_0 = \SI{2}{\micrometer}$. (\textbf{c-d}) Joint radial distribution $P(\Delta r_{ij} , \Delta t_{r,ij} ) / \Delta r_{ij}$ of spatial and temporal separations among pairs of reorientation events in the null model, which consists of randomizing the angular positions of cells within the biofilm, for (\textbf{c}) $\ell_0 = \SI{1}{\micrometer}$ and (\textbf{d}) $\ell_0 = \SI{2}{\micrometer}$.
}
\end{figure}

To rule out that the peak structure in the joint radial distribution of verticalization events was caused by the finite size and growth rate of the annular region, we compared our results to a null model that accounts for this effect by randomizing the angular positions of cells within the biofilm. Specifically, for the null model, we compute the spatiotemporal separations between verticalization events from a given biofilm to those in ten copies of the same biofilm that have been randomly rotated around its center. This model respects the radial symmetry of the biofilm and also allows for comparison between biofilms of different average cell lengths. Specifically, if correlations within a given biofilm exceed those obtained for the corresponding null model, then any excess correlation implies a nontrivial source of correlations. We found that the probability at the peak was significantly increased compared to the null model, which demonstrates that verticalization events are cooperative. This effect is similar to the phenomenon of dynamical facilitation observed in glassy systems\cite{S4}. A possible explanation for the nontrivial correlations comes from the inverse domino effect, which consists of a cell standing up and applying an out-of-plane torque that triggers one or more neighboring cells to stand up. We expect this effect to occur for long cells because long cells are more likely to become vertical due to the peeling instability, which is triggered by torques from neighboring vertical cells.

Does the inverse domino effect explain the spatiotemporal extent of the peak? The inverse domino effect can occur when a vertical cell comes into contact with a horizontal cell that is susceptible to becoming vertical. The requirement of cell-cell contact for this effect to occur is consistent with the observed spatial separation of the peak, which is approximately equal to the average distance between the centers of horizontal cells (Supplementary Fig. 8). Furthermore, the inverse domino effect also suggests a limit on the temporal extent of the peak, because the reduction of cell footprint upon verticalization opens up space for local rearrangements that rapidly alleviate the surface pressure as the cell configuration relaxes (Supplementary Fig. 5). Indeed, the time it takes for the surface pressure to relax is roughly a few minutes, consistent with the temporal extent of the peak beyond its maximum (Supplementary Figs. 5, 8). Thus, taken together, the requirement for spatial proximity, along with the decrease in surface pressure associated with verticalization, can explain the rapid spatiotemporal decay of $P(\Delta r_{ij} , \Delta t_{r,ij} )$.

Our observations of the behavior of $P(\Delta r_{ij} , \Delta t_{{r},ij} )$ provide a natural definition for the extent of cooperativity in cell verticalization. That is, since facilitation occurs on short spatiotemporal scales, we can capture the extent of cooperative effects by computing the number of cells involved in a series of verticalization events that are proximal in space and time. Specifically, we define proximity in space as $\Delta r_{ij} < \ell_{{f}}$, where $\ell_{{f}}$ is the cell division length, and define proximity in time as $\Delta t_{{r},ij} < t_{{f}}$, where $t_{{f}}$ is the facilitation time scale, defined as the time scale of the decay of the spatiotemporal separation probability $P(\Delta r_{ij} , \Delta t_{r,ij} )/ \Delta r_{ij}$ after the peak. For the growth of a given biofilm cluster, connecting reorientation events that are spatiotemporally proximal results in a graph. We refer to the connected components of this graph as ``cell avalanches'', following the literature on glasses\cite{S4}.

Interestingly, the distribution of avalanche sizes decays roughly exponentially for all values of cell length we studied (Supplementary Fig. 9), with only a modest number of cells involved in typical avalanches (Fig. 2d). Moreover, the distribution of avalanche sizes does not change substantially as a function of time, unlike the overall number of horizontal cells, which grows proportionally to the biofilm radius (Fig. 1). Thus, as time goes on, a vanishing fraction of the overall number of horizontal cells are involved in a typical avalanche, which demonstrates that avalanches are localized. The dependence of the mean avalanche size on cell length demonstrates that the scale of localization is determined by the geometrical and mechanical properties of individual cells (Fig. 2d).

What limits the size of cell avalanches? In order to be susceptible to becoming vertical due to the inverse domino effect, horizontal cells must be poised near the threshold torque for verticalization. Thus, a natural explanation for the size limit comes from the reduction of cell footprint upon reorientation from horizontal to vertical, which rapidly alleviates the local surface pressure and thereby lowers the susceptibility of nearby horizontal cells to becoming vertical (Supplementary Fig. 5). This effect combines with the inherent disorder in the configuration of cells, which generically results in extremely heterogeneous contact geometries and forces throughout the biofilm (Supplementary Video 3, Supplementary Fig. 10). These heterogeneous local conditions segregate horizontal cells poised near the threshold torque for verticalization into small groups. Although within such groups, the verticalization cooperativity is transiently increased by the inverse domino effect, verticalization rapidly exhausts the local supply of horizontal cells. Thus, the rapid timescale of verticalization and the disorder in the cell configuration limit the propagation of cell avalanches.

\newpage

\begin{figure}[H]
\centering
\includegraphics[width=0.8\columnwidth]{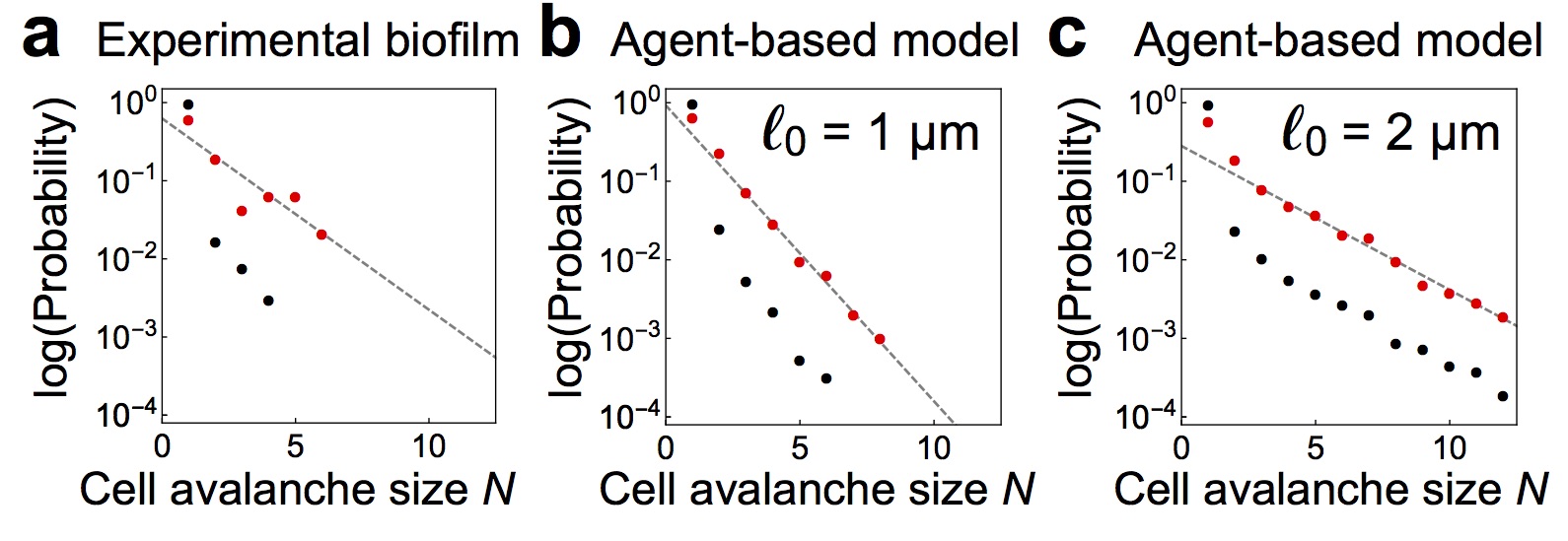}
\caption{\label{fig:FG7}
Representative examples of avalanche size distributions. (\textbf{a-c}) Distributions of avalanche sizes $N$ (red data points), defined as the number of cells in a group of verticalization events that are proximal in space (i.e. separated by a distance $\Delta r_{ij} < \ell_{{f}}$, where $\ell_{{f}}$ is the cell division length) and time (i.e. with temporal separation $\Delta t_{{r},ij} < t_{{f}}$) on a logarithmic scale for (\textbf{a}) experimental biofilm, (\textbf{b}) agent-based model with $\ell_0 = \SI{1}{\micrometer}$, and (\textbf{c}) agent-based model with $\ell_0 = \SI{2}{\micrometer}$. The black data points indicate the corresponding distribution of avalanche sizes for a null model. For reference, gray straight dashed lines correspond to exponential decay over a scale (\textbf{a}) $N=1.8$ cells, (\textbf{b}) $N=1.2$ cells, and (\textbf{c}) $N=2.4$ cells.
}
\end{figure}

\begin{figure}[H]
\centering
\includegraphics[width=0.8\columnwidth]{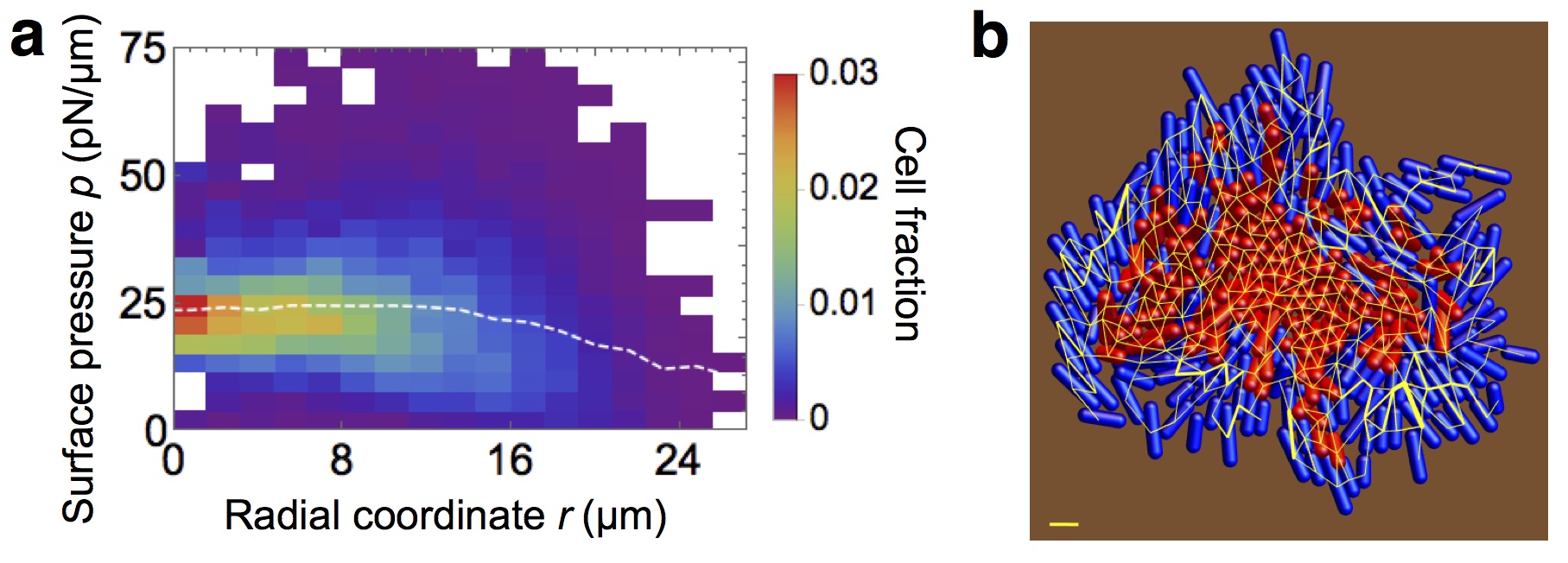}
\caption{\label{fig:FG9005}
Mechanical heterogeneity in the agent-based model. (\textbf{a}) Joint distribution of cell surface pressures $p$, defined as the sum of all horizontal forces acting on a cell divided by its perimeter, and radial coordinates $r$ for cells in modeled biofilm with $\ell_0 = \SI{1.2}{\micrometer}$, showing cell fraction in color. Dashed white curve shows the average cell surface pressure $\langle p \rangle$ versus $r$. (\textbf{b}) Visualization of surface layer of a modeled biofilm with $\ell_0 = \SI{2}{\micrometer}$, showing horizontal (blue) and vertical (red) cells as spherocylinders, the surface (brown), and cell-to-cell contact forces (yellow lines connecting the centers of cells, with thickness proportional to the force). Cells with $n_z < 0.5$ ($>0.5$) are considered horizontal (vertical), where $\boldsymbol{\hat{n}}$ is the orientation vector. The length of the scale bar is $\SI{3}{\micrometer}$, and its thickness corresponds to $\SI{300}{\pico\newton}$.
}
\end{figure}

\newpage

\section*{\textbf{Supplementary Note: Continuum models for verticalizing biofilms}}

In this Supplementary Note, we present minimal continuum models that provide insight into the verticalization transition. We first present a simplified model for verticalizing biofilms in the incompressible limit. We go on to explore the compressible biofilm model that was discussed in the main text.

\subsection*{\textbf{Origin of vertical ordering}}

How does cell growth drive biofilm expansion and verticalization? To gain qualitative insight into this question, we started by considering a simple continuum model in the limit of approximately incompressible cells. We first assumed that cells in 2D grow exponentially at rate $\alpha$. For an isotropic 2D biofilm, this growth implies that the total radius $R_{{B}}$ increases as:

\begin{equation}
R_{{B}} \sim e^{\alpha t / 2}.
\end{equation}

\noindent Similarly, the local radial velocity must be $v = \alpha r / 2$, where $r$ is the radial coordinate. This velocity must arise from the cell surface pressure $p$, associated with the compression of cells. The local gradient of this surface pressure $dp/dr$ required to drive cells with velocity $v$ is

\begin{equation}
\frac{dp}{dr} = -\eta v = - \frac{\alpha \eta  r }{2},
\end{equation}

\noindent where $\eta$ is the surface drag coefficient of the medium. Spatially integrating this equation gives the pressure field:

\begin{equation}
p = \frac{\alpha \eta}{4} (R_{{B}}^2 - r^2),
\end{equation}

\noindent which is quadratic and peaked at the center of the biofilm.

Now we assume that as soon as the local surface pressure exceeds some verticalization threshold $p_{t}$, the cells start becoming vertical. These transitions occur first at the center of the biofilm, resulting in an inner region of vertical cells surrounded by an annular periphery of horizontal cells. Since vertical cells do not contribute to growth along the surface, the surface pressure remains constant throughout the region of vertical cells. Furthermore, to satisfy the boundary condition $p=p_{t}$ at the interface between horizontal and vertical cells, the width of the annular periphery of horizontal cells must remain constant. This results in a biofilm front and a region of vertical cells that both expand outward at a fixed rate $c^* \sim \sqrt{\alpha  p_{t}/ \eta }$. Thus, this simple continuum model of incompressible cells provides a qualitative explanation of the verticalization transition.

Although this model roughly captures the spreading of vertical ordering, it cannot capture the crossover between the radial density profiles of the horizontal and vertical cells (Fig. 1c,d), or the saturation of the expansion speed as a function of increasing verticalization threshold (Results, Equation 2).

\subsection*{\textbf{Compressible, two-fluid model for verticalizing biofilms}}

To better quantify the growth of verticalizing biofilms, we developed a continuum model that treats horizontal and vertical cells, respectively, as densities $\rho_h$ and $\rho_v$. These densities specify the number of cells per unit of surface area. In what follows, we define the total cell density as $\tilde{\rho}_{\mathrm{tot}} = \rho_h + \xi \rho_v$, where $\xi$ is the ratio of vertical to horizontal cell footprints. The cell densities evolve according to the following hydrodynamic equations:

\begin{equation}
\dot{\rho_h} + \nabla \cdot ( \rho_h \boldsymbol{v}) = \left[ \alpha - \beta \Theta(p - p_t) \right] \rho_h,
\end{equation}

\begin{equation}
\dot{\rho_v} =  \beta \Theta(p - p_t) \rho_h,
\end{equation}

\begin{equation}
- \eta \tilde{\rho}_{\mathrm{tot}} \boldsymbol{v} = \nabla p.
\end{equation}

\noindent Here, $\boldsymbol{v}$ is the cell velocity, $\alpha$ is the growth rate, $\beta$ is the verticalization rate, $\eta$ is a viscous drag coefficient, $p$ is the surface pressure, $p_t$ is the threshold surface pressure for verticalization, $\tilde{\rho}_0$ is the close-packing density, and $\Theta$ is the Heaviside step function. The first two equations describe the conservation of cell number, and the third equation describes the balance between growth forces and surface drag. In the first equation, we have assumed that the change in local horizontal cell density is determined by the effect of cell transport, i.e. the change in cell density due to the motion of cells, as well as in-plane cell growth and cell verticalization. In the second equation, we have neglected the transport of vertical cells, and so the vertical cell density only changes due to cell verticalization. Below, we will revisit this approximation and show that it does not change the results for rapid enough verticalization rates $\beta > \alpha(1-\xi)$ consistent with the behavior of the experimental biofilms and the agent-based modeled biofilms (Fig. 4b).

We take the surface pressure to be given by the Young's modulus $\lambda$ of the biofilm times the areal strain, which becomes nonzero when cells are close-packed but uncompressed:

\begin{equation} 
p=\begin{cases}
    \lambda ( \tilde{\rho}_{\mathrm{tot}} - \tilde{\rho}_0) & \text{if}\  \tilde{\rho}_{\mathrm{tot}} > \tilde{\rho}_0,\\
    0              & \text{otherwise}.
\end{cases}
\end{equation}

\noindent Therefore, in our model, the threshold surface pressure for verticalization corresponds to a threshold surface density $\tilde{\rho}_t$ for verticalization:

\begin{equation}
p_t = \lambda (\tilde{\rho}_t - \tilde{\rho}_0).
\end{equation}

Upon substituting this relation between the pressure and cell density into the hydrodynamic equations, we obtain the following equations of motion for the cell densities:

\begin{equation}
\dot{\rho_h} =  \frac{\lambda}{\eta} \nabla \cdot \left(\Theta(\tilde{\rho}_{\mathrm{tot}} - \tilde{\rho}_0)  \nabla \tilde{\rho}_{\mathrm{tot}}\right)  +  \left[ \alpha - \beta \Theta(p - p_t) \right] \rho_h,
\end{equation}

\begin{equation}
\dot{\rho_v}  =  \beta \Theta(p - p_t) \rho_h.
\end{equation}

In the following sections, we solve for the dynamics of the cell densities. For simplicity, we will first solve the case of spreading in one spatial dimension, which will allow us to characterize the different dynamical regimes of the model. In the last few sections, we will discuss how the results are modified for the cases of non-stationary vertical cells, two dimensional growth, and surface curvature.

\subsection*{\textbf{Existence of a linearly-expanding front}}

To understand whether our continuum model can give rise to stable, linearly-expanding fronts, we first consider the growth of the total number of horizontal cells throughout the entire biofilm. The change in the number of horizontal cells is given by:

\begin{equation}
\dot{\rho}_h = \int_{\mathcal{R}_1} \alpha \rho_h d\boldsymbol{r} + \int_{\mathcal{R}_2} (\alpha - \beta) \rho_h d\boldsymbol{r},
\end{equation}

\noindent where $\mathcal{R}_1$ corresponds to regions with $p<p_t$ and $\mathcal{R}_2$ corresponds to regions with $p>p_t$. For $\alpha > \beta$, both terms are positive, and the total number of horizontal cells must grow exponentially with rate $\alpha - \beta$ at long times. At long times, the regions $\mathcal{R}_2$ will dominate the growth. Assuming a uniform growth rate $\alpha - \beta$, the radius $R_B$ of the biofilm is given by:

\begin{equation}
R_B \sim \sqrt{(\alpha - \beta) \gamma ( t^2 - t \log t )},
\end{equation}

\noindent at long times\cite{S5}, which yields an expansion speed $c^*$ given by:

\begin{equation}
c^* \sim \sqrt{(\alpha-\beta) \gamma} \left(1 - \frac{1}{t}\right).
\end{equation}

Thus, the speed of the edge of the biofilm cluster increases over time. However, for $\beta>\alpha$, the contribution from region $\mathcal{R}_2$ is negative and thus could potentially compensate the contribution from region $\mathcal{R}_1$ to limit the total growth of horizontal cells and thereby allow for a linearly-expanding front.

\subsection*{\textbf{Solving for steady-state motion}}

To search for linearly-expanding solutions to the equations that govern the dynamics of the cell densities, we now assume that the biofilm expands linearly with a speed $c^*$ and we seek consistent solutions to the equations of motion. For now, we treat $c^*$ as an undetermined constant. To solve the equations of motion, we start by shifting to a reference frame that moves at speed $c^*$:

\begin{equation}
0 =  \frac{\lambda}{\eta} \Theta(\tilde{\rho}_{\mathrm{tot}} - \tilde{\rho}_0) \nabla^2\tilde{\rho}_{\mathrm{tot}}  + c^* \nabla \rho_h + \left[ \alpha - \beta \Theta(p - p_t) \right] \rho_h,
\end{equation}

\begin{equation}
0  =  c^* \nabla \rho_v + \beta \Theta(p - p_t) \rho_h.
\end{equation}

\noindent We can eliminate the density of vertical cells from the first equation by substituting in the second equation. Doing this yields the following equation:

\begin{equation}
0 =  \frac{\lambda}{\eta}  \Theta(\rho_h + \xi \rho_v - \tilde{\rho}_0) \nabla^2 \rho_h  + \left( c^* - \frac{\lambda \beta \xi}{\eta c^*}\Theta(p - p_t) \right) \nabla \rho_h + \left[ \alpha - \beta \Theta(p - p_t) \right] \rho_h,
\end{equation}

To be consistent with the biofilm morphology observed in experiment, we assume that the leading edge of the biofilm consists of a periphery of horizontal cells trailed by an interior region containing a mixture of horizontal and vertical cells with $p>p_t$. For the continuum model, we verified, using numerical simulations, that this pattern generically arises from initial conditions that consist of a small, localized region of horizontal cells (Methods, Supplementary Video 4). Below, we will determine the conditions under which these preliminary solutions are valid. These assumptions lead to the following equations for the horizontal periphery (``$P$'') and the mixed interior (``$I$''):

\begin{equation}
0 =   \frac{\lambda}{\eta} \nabla^2 \rho_h^{(P)}   + c^* \nabla \rho_h^{(P)} + \alpha  \rho_h^{(P)},
\end{equation}

\begin{equation}
0 =   \frac{\lambda}{\eta} \nabla^2 \rho_h^{(I)}  + \left( c^* - \frac{\lambda \beta \xi}{\eta c^*}  \right) \nabla \rho_h^{(I)}  +(\alpha - \beta) \rho_h^{(I)}.
\end{equation}

\subsection*{\textbf{Boundary conditions}}

At the leading edge of the horizontal periphery, we must have:

\begin{equation}
\rho_h^{(P)} = \tilde{\rho}_0,
\end{equation}

\noindent since the pressure that drives cell motion drops to zero when the cell density declines below the packing density $\tilde{\rho}_0$. Furthermore, the leading edge of the horizontal periphery must be moving at speed $c^*$:

\begin{equation}
- \rho_h^{(P)}c^* = \frac{\lambda}{\eta} \nabla \rho_h^{(P)}.
\end{equation}

\noindent The interface between the horizontal periphery and the mixed interior marks the onset of verticalization, which implies:

\begin{equation}
\rho_h^{(I)} = \rho_h^{(P)} = \tilde{\rho}_t.
\end{equation}

Finally, the surface pressure gradient must be continuous at the interface:

\begin{equation}
\nabla \rho_h^{(I)} + \xi \nabla \rho_v^{(I)} = \nabla \rho_h^{(P)},
\end{equation}

\subsection*{\textbf{The density of horizontal cells in the mixed interior must be given by a single, exponentially-decaying term}}

From the above equation of motion for the mixed interior, we find:

\begin{equation}
\rho_h^{(I)} = q_1 e^{\gamma_+ \tilde{x}} + q_2 e^{\gamma_- \tilde{x}},
\end{equation}

\noindent where $q_1$ and $q_2$ are constants to be determined by the boundary conditions,

\begin{equation}
\gamma_{+,-} = \left( \frac{\beta \xi}{2c^*} - \frac{\eta c^*}{2 \lambda}  \right) \pm \sqrt{ \left(\frac{\beta \xi}{2c^*} - \frac{\eta c^*}{2 \lambda} \right)^2 - \frac{\eta}{\lambda}(\alpha - \beta)},
\end{equation}

\noindent and we have chosen the spatial coordinate $\tilde{x}$ such that $\tilde{x} = 0$ is the location of the interface between the two regions. We can further simplify this by inserting the boundary condition for the cell density at the interface to eliminate one of the undetermined constants. We find:

\begin{equation}
\rho_h^{(I)} = q_1 e^{\gamma_+ \tilde{x}} + (\rho_t - q_1)  e^{\gamma_- \tilde{x}}.
\end{equation}

For $\alpha < \beta$, $\gamma_+$ and $\gamma_-$ must both be purely real. Furthermore, since $  | \frac{\beta \xi}{2c^*} - \frac{\eta c^*}{2 \lambda} | <  \sqrt{ \left( \frac{\beta \xi}{2c^*} - \frac{\eta c^*}{2 \lambda} \right)^2 -  \frac{\eta}{\lambda}(\alpha - \beta) }$, the constants must have opposite signs. That is, $\gamma_+$ is positive and $\gamma_-$ is negative.

We now show that if the density of horizontal cells is 0 at the inner boundary, the density of horizontal cells must be given by a decaying exponential. Consider a density of horizontal cells that is $\rho_t$ at $\tilde{x}=0$ and approaches zero at some finite negative value $\tilde{x}=\tilde{x}_t$. Since the density of horizontal cells at $\tilde{x}=\tilde{x}_t$ is zero, the cell flux must also be zero (since $-\eta \rho_h \boldsymbol{v}_h = \lambda \nabla \rho_h$). Therefore, at this inner boundary, we have:

\begin{equation}
q_1 e^{\gamma_+ \tilde{x}_t } + (\tilde{\rho}_t - q_1) e^{\gamma_- \tilde{x}_t } = 0,
\end{equation}

\begin{equation}
\gamma_+ q_1 e^{\gamma_+ \tilde{x}_t } - \gamma_- (\tilde{\rho}_t - q_1) e^{\gamma_- \tilde{x}_t } = 0.
\end{equation}

\noindent These equations only have a non-trivial solution for $b_1 = 0$, which means that the density of horizontal cells must be given by an exponential function with a decay constant $\gamma_+$:

\begin{equation} \label{eq:hord}
\rho_h^{(I)} =  \tilde{\rho}_t e^{\gamma_+ \tilde{x} }.
\end{equation}

\subsection*{\textbf{Solving for the expansion speed}}

We now determine the expansion speed $c^*$ by solving for the steady-state density of horizontal cells at the leading edge of the biofilm. We choose coordinates $x$ such that the front of this leading edge is at $x=0$. Upon insertion of the boundary conditions at the front of the leading edge, we find that:

\begin{equation}
\rho_h^{(P)} = \tilde{\rho}_0 e^{-\frac{\eta c^* x}{2 \lambda}} \left[ \cosh \left(\frac{\eta c^* x}{2\lambda}  \sqrt{ 1 - \frac{4 \alpha \lambda}{\eta c^{*2}} }\right)-\frac{\sinh \left(\frac{\eta c^* x}{2\lambda}  \sqrt{ 1 - \frac{4 \alpha \lambda}{\eta c^{*2}} } \right)}{  \sqrt{ 1 - \frac{4 \alpha \lambda}{\eta c^{*2}}  }}\right].
\end{equation}

This profile extends to negative values of $x$ until the density of horizontal cells reaches $\rho_h = \tilde{\rho}_t$ at the interface $x=x_t$. The value of $x_t$ may be obtained from the following non-dimensionalized form of the above equation:

\begin{equation}
\frac{\tilde{\rho}_t}{\tilde{\rho}_0} = e^{-\kappa q} \left(\cosh(\kappa q)- \frac{\sinh(\kappa q)}{\kappa} \right),
\end{equation}

\noindent where $q = \frac{\eta c^*}{2 \lambda} x_t$ and $\kappa = \sqrt{1 - \frac{4 \alpha \lambda } {\eta c^{*2}} }$. Finally, at the interface, we insert the boundary condition for the balance of cell flux, which states:

\begin{equation}
\tilde{\rho}_t \left( \gamma_+  - \frac{\xi \beta}{c^*} \right) = \nabla \rho_h^{(P)}(x=x_t).
\end{equation}

This equation has the following non-dimensional form:

\begin{equation}
\frac{\tilde{\rho}_t}{\tilde{\rho}_0} = \frac{ e^{-w} \sinh ({w^2 - \delta^2}) }{\sqrt{w^2 - \delta^2}} \left( \frac{\xi \chi \delta}{4w} - \frac{w}{\delta} - \sqrt{ \left(\frac{w}{\delta} + \frac{\xi \chi \delta}{4w} \right)^2 - \delta \left[ 1 + \chi (\xi - 1) \right] } \right)^{-1},
\end{equation}

\noindent where $\delta^2= \alpha \eta x_t^2 / \lambda$, $w = \eta x_t v / 2 \lambda$, and $\chi = \beta / \alpha$. This equation determines the speed of the expanding front, provided our assumption holds that the surface pressure exceeds the threshold surface pressure for reorientation in the mixed interior.

\subsection*{\textbf{The assumption of simple verticalization in the mixed interior can break down}}

For the above solution to be consistent, the surface pressure in the mixed interior region must always exceed the threshold surface density $\tilde{\rho}_{t}$ for reorientation:

\begin{equation}
\rho_h + \xi \rho_v > \tilde{\rho}_{t}.
\end{equation}

\noindent The density $\rho_v$ of vertical cells is obtained by integrating the density of horizontal cells:

\begin{equation}
\rho_v = -\frac{\beta}{c^*} \int_0^x \rho_h dx.
\end{equation}

\noindent Inserting the solution for $\rho_h$ from above (see Supplementary Eq. \ref{eq:hord}), we find that $\rho_v$ is given by:

\begin{equation}
\rho_v = \frac{\beta}{c^* \gamma_+} (1 - e^{\gamma_+ x}). 
\end{equation}

\noindent The condition for the surface density to exceed the threshold surface density is:

\begin{equation}
\tilde{\rho}_t \left( \frac{\xi \beta}{c^* \gamma_+} - \frac{\xi \beta}{c^* \gamma_+} e^{\gamma_+ x} + e^{\gamma_+ x} \right) > \tilde{\rho}_t.
\end{equation}

\noindent The slope of the surface density is given by $(\gamma_+ - \xi \beta / c^* )e^{\gamma_+ x}$. The exponential part is always positive, and its prefactor is:

\begin{equation}
(\gamma_+ - \xi \beta / c^* ) = \frac{-(\xi \beta + c^{*2}) + \sqrt{(\xi \beta + c^{*2})^2 - 4 \frac{\lambda}{\eta} c^{*2} (\alpha + \xi \beta - \beta)}}{2c^*(\lambda/\eta)}.
\end{equation}

\noindent For large enough ratio of growth rate to verticalization rate, i.e. $\alpha / \beta > 1 - \xi$, this quantity is always negative since $\xi \beta + c^{*2} > \sqrt{(\xi \beta + c^{*2})^2 - 4 \frac{\lambda}{\eta} c^{*2} (\alpha + \xi \beta - \beta) }$. In this case, the surface density increases monotonically as $x\rightarrow -\infty$, and the surface density always exceeds the threshold surface density for verticalization in the mixed interior.

\subsection*{\textbf{For high verticalization rates, dynamical isobaricity determines the cell density profiles}}

In the previous section, we found that our candidate solution for the steady-state cell density in the mixed interior could yield a surface density profile that was too small to sustain the assumed verticalization. Specifically, the solution fails when $\alpha / \beta < 1 - \xi$. Intuitively, this occurs because the horizontal cells become vertical faster than the maximum rate at which the combined effects of cell growth and cell transport can replenish the threshold surface density of cells needed for further verticalization to occur. Thus, when $\alpha / \beta < 1-\xi$, the surface density must constantly fluctuate between verticalizing and non-verticalizing values. This fluctuation stabilizes the surface density throughout the mixed interior at the verticalization threshold. The stability of the uniform surface density is apparent from examining the equation for the surface pressure $p \sim \rho_h + \xi \rho_v - \tilde{\rho}_0$:

\begin{equation}
\dot{p}= \frac{\lambda}{\eta} p'' + c^* p' + \alpha p + \left( \xi \beta - \beta \right) \Theta(p - p_t) p.
\end{equation}

\noindent Consider a distribution of cells throughout the mixed interior such that the surface pressure is everywhere equal to the threshold surface pressure for verticalization. Here, a verticalization event may bring the total surface pressure below the threshold surface pressure at some particular location. At such a location, the total surface pressure is at a local minimum, which means that $p' = 0$ and $p'' > 0$. Therefore the rate of change of surface pressure is positive:

\begin{equation}
\dot{p} =  \frac{\lambda}{\eta} p'' + \alpha p > 0.
\end{equation}

\noindent Conversely, if the total surface pressure ever exceeds the threshold surface pressure at some location, we have $p' = 0$ and $p'' < 0$, and the rate of change in the surface pressure must be negative:

\begin{equation}
\dot{p}=  \frac{\lambda}{\eta} p'' +\alpha p + \left( \xi \beta - \beta \right) \Theta(p - p_t) p.
\end{equation}

\noindent Therefore, in the regime of rapid verticalization, any deviation of the total surface pressure away from the threshold surface pressure will decay in time. This argument suggests that at any specific location in the mixed interior, the surface pressure will be close to the threshold surface pressure, and the horizontal cells will constantly fluctuate between verticalizable and non-verticalizable conditions. To predict the cell density profile in the mixed interior, we assume that the surface pressure is maintained at the threshold surface pressure by cells that spend a fraction of time $\kappa$ in the verticalizable state. This results in the following equations of ``steady state'':

\begin{equation}
0 = c^* \rho_h' + \left[ (\alpha - \beta)\kappa + \alpha(1 - \kappa) \right] \rho_h,
\end{equation}

\begin{equation}
0 = c^* \rho_v' + \beta \kappa \rho_v.
\end{equation}

Since the surface pressure is constant, we have $\tilde{\rho}_t = \rho_h + \xi \rho_v$, which implies:

\begin{equation}
\rho_h' = -\xi \rho_v',
\end{equation}

\noindent which, together with the above equations, allows us to solve for the fraction of time $\kappa$ spent in the verticalizable state:

\begin{equation}
\kappa = \frac{\alpha}{\beta (1 - \xi)}.
\end{equation}

Inserting this into the equations of steady state yields horizontal and vertical cell densities that decay and grow exponentially, respectively, at a rate $\mu$ given by:

\begin{equation} \label{eq:dec}
\mu = \frac{\alpha \xi }{c^*(1-\xi)},
\end{equation}

\noindent which, interestingly, does not depend on the verticalization rate $\beta$. The resulting horizontal and vertical cell surface density profiles determine the boundary conditions at the interface between the mixed interior and the horizontal cell periphery, which thereby determines the overall expansion speed of the biofilm.

\subsection*{\textbf{Phase diagram for verticalizing biofilms}}

We summarize our results from the continuum modeling in the phase diagram in Supplementary Fig. 11\textbf{a}. The dynamics of the cell densities fall into three different regimes, depending on the values of the growth rate $\alpha$, the verticalization rate $\beta$, and the ratio $\xi$ of vertical to horizontal cell footprints. For $\alpha>\beta$, the overall number of horizontal cells increases exponentially, which precludes the existence of a stable, linearly-propagating front. For $\alpha < \beta$, the biofilm develops into a mixed interior of vertical and horizontal cells surrounded by a periphery of horizontal cells, which both spread outwards linearly in time. For $\beta (1-\xi) < \alpha < \beta$, the surface pressure and density continue to build up inside the mixed interior and ultimately saturate at values above the threshold values for verticalization. However, when $\beta (1-\xi) < \alpha$, verticalization can deplete the cell density in the mixed interior more rapidly than cell density can be replenished by cell growth and cell transport due to gradients in surface pressure. Thus, in this regime, the surface pressure and density rapidly fluctuate around the threshold values for verticalization, which effectively tunes the verticalization rate to $\beta (1-\xi) = \alpha$.

\subsection*{\textbf{The effect of vertical cell transport}}

Incorporating an in-plane velocity for vertical cells into the equations for the change in cell densities yields:

\begin{equation}
\dot{\rho_h} + \nabla \cdot ( \rho_h \boldsymbol{v}) = \left[ \alpha - \beta \Theta(p - p_t) \right] \rho_h,
\end{equation}

\begin{equation}
\dot{\rho_v} + \nabla \cdot ( \rho_v \boldsymbol{v})=  \beta \Theta(p - p_t) \rho_h,
\end{equation}

\begin{equation}
- \eta \tilde{\rho}_{\mathrm{tot}} \boldsymbol{v} = \nabla p.
\end{equation}

The presence of vertical cell transport could potentially influence the dynamics in regions with a finite vertical cell density, i.e. the mixed interior. For the isobaric regime $\alpha < \beta(1-\xi)$, the vertical cell transport has no effect because there the cell velocity is zero in the mixed interior due to the uniformity of the surface pressure. Furthermore, for $\alpha < \beta$, the same argument given above (see \textbf{Existence of a linearly-expanding front}) implies that there is no stable, linearly-expanding steady state. To examine how vertical cell transport impacts the non-isobaric regime, we incorporated this effect into the simulations. We found that the presence of vertical cell transport does not qualitatively change the cell density profiles. Furthermore, incorporating vertical cell transport only results in small changes to the cell densities of around a few percent (inset, Supplementary Fig. 11b). Thus, although the presence of vertical cell transport can slightly affect the quantitative details of the cell density profiles in the non-isobaric regime, vertical cell transport does not influence the qualitative behaviors of the different phases of our model.

\subsection*{\textbf{Growth in two spatial dimensions}}

In two spatial dimensions, the equation for the total cell density (assuming no vertical cell transport) becomes:

\begin{equation}
\dot{\tilde{\rho}}_{\mathrm{tot}} = \gamma \left(\frac{1}{r}\frac{d}{dr}\tilde{\rho}_{\mathrm{tot}} + \frac{d^2}{dr^2}\tilde{\rho}_{\mathrm{tot}} \right) + \alpha  \rho_h - (1 - \xi) \beta \Theta(\tilde{\rho}_{\mathrm{tot}} - \tilde{\rho}_t) \rho_h.
\end{equation}

The effect of the additional spatial dimension is to add an advection-like term $\frac{1}{r}\frac{d}{dr}\tilde{\rho}_{\mathrm{tot}}$ to the change in cell density, which, as above, does not alter the qualitative features of the model's phase behavior. Furthermore, this term becomes less important for larger values of the radial coordinate $r$. To estimate the importance of this term, we compare $r$ to the maximum value of $\frac{d}{dr}\tilde{\rho}_{\mathrm{tot}}$, which is given by $\frac{d}{dr}\tilde{\rho}_{\mathrm{tot}}=\rho_0 c^* / \gamma$ at the edge of the biofilm. This estimate suggests that for large values of the radial coordinate $r \gg \rho_0 c^* / \gamma$, the dynamics become effectively one-dimensional.

\subsection*{\textbf{The effect of surface curvature}}

How does surface curvature influence the global build-up of surface pressure? Here, we answer this question by extending our incompressible model (see \textbf{Origin of vertical ordering}) to the case of a curved surface. Specifically, we consider an incompressible biofilm growing exponentially at a rate $\alpha$ along a sphere of radius $R$. We assume that growth begins from a single point and that the biofilm remains azimuthally symmetric. Thus, the extent of the biofilm along the surface, defined as the geodesic distance along the surface between the origin $\mathcal{O}$ of growth and the edge of the biofilm, is described by a single coordinate $R_B$. For a given value of $R_B$, the surface area covered by the biofilm is given by:

\begin{equation}
A = 2 \pi R^2 \left( 1 - \cos \left( \frac{R_B}{R} \right) \right).
\end{equation}

The exponential growth of $A$ implies that the velocity $v = dR_B / dt$ of the biofilm extent is given by:

\begin{equation}
v = \frac{\alpha A}{2\pi \sqrt{1 - R \left( 1 - \frac{A}{2\pi R^2} \right)^2}}.
\end{equation}

Similarly, the local velocity of a point at a distance $r$ along the surface from $\mathcal{O}$ is given by the same expression, but with $A$ instead giving the area of the spherical cap that contains all points nearer to the origin of growth than $r$. As before for growth on a flat surface, we assume that this velocity is driven by the local gradient $dp/dr$ of surface pressure:

\begin{equation}
\frac{dp}{dr} = -\eta v,
\end{equation}

\noindent where $\eta$ is the surface drag coefficient of the cell medium. Spatially integrating this equation gives the pressure field:

\begin{equation}
p = 2 \eta \alpha R^2 \left[ \log \cos \left( \frac{r}{2R} \right)  - \log \cos \left( \frac{R_B}{2R} \right)   \right],
\end{equation}

\noindent which is peaked at the origin of growth.

This equation indicates that, for a given extent $R_B$, the effect of surface curvature is to increase the surface pressure throughout the biofilm. Intuitively, this increase arises because, for a fixed biofilm extent $R_B$, the biofilm footprint is larger for a flat surface than for a spherical surface. This difference in footprint implies that, assuming equal growth rates, cells must spread out more rapidly on the spherical surface to accommodate the increase in total biofilm surface area. In turn, the increase in biofilm expansion speed implies a larger gradient in surface pressure. This argument suggests that negatively-curved surfaces, e.g. saddle-like surfaces, would have the opposite effect; they would decrease the rate at which surface pressure builds up.

For values of $r \ll R$, the surface of the sphere is effectively flat on the scale of the biofilm, and the equation for the surface pressure on a sphere reduces to the expression for a flat surface. However, when the biofilm extent becomes comparable to the radius of curvature of the surface, the increase in surface pressure becomes substantial (Supplementary Fig. 14). This increase in surface pressure can trigger verticalization at much smaller values of $R_B$ for a spherical surface than for a flat surface. Thus, surface curvature provides an additional geometrical mechanism that regulates the transition of biofilms to three-dimensional growth.

\newpage

\section*{\textbf{Supplementary Figure 11: Phase diagram for verticalizing biofilms}}

\begin{figure}[H]
\centering
\includegraphics[width=0.8\columnwidth]{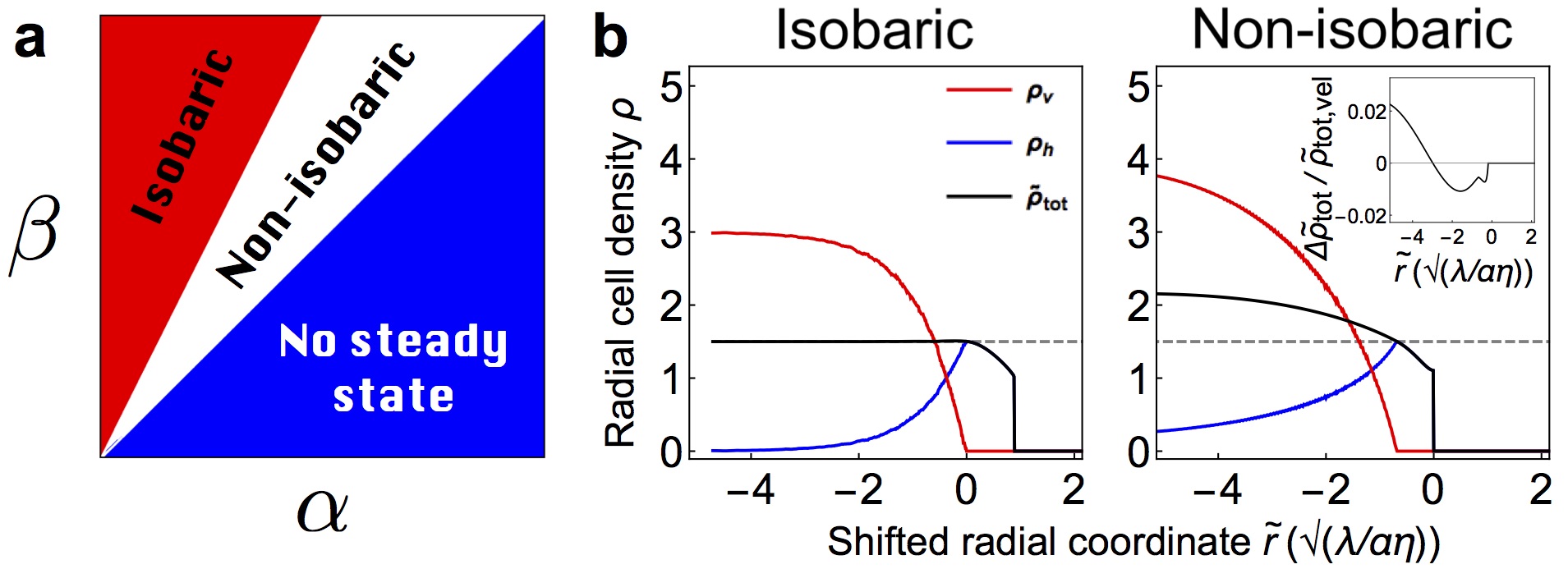}
\caption{\label{fig:FG9019}
Phases of front propagation of verticalizing biofilms. (\textbf{a}) Phase diagram of continuum model for verticalizing biofilms. For $\alpha < \beta(1-\xi)$, the mixed interior is isobaric, i.e. the surface pressure is uniform. For $\beta  >\alpha > \beta(1-\xi)$, the surface pressure of the mixed interior decreases monotonically with the radial distance from the biofilm center. For $\alpha > \beta$, there is no stable, steady-state linearly-propagating front. (\textbf{b}) Numerical simulations of the continuum model assuming no vertical cell transport, showing radial densities of horizontal cells ($\rho_h$, blue), vertical cells ($\rho_v$, red), and total density ($\tilde{\rho}_{\mathrm{tot}}$, black), versus shifted radial coordinate $\tilde{r}$ for isobaric regime with $\tilde{\rho}_0 = \SI{1}{\meter^{-2}}$, $\tilde{\rho}_t = \SI{1.5}{\meter^{-2}}$, $\beta = 2.5 \alpha$, and $\xi=0.5$ (left) and non-isobaric regime with $\tilde{\rho}_0 = \SI{1}{\meter^{-2}}$, $\tilde{\rho}_t = \SI{1.5}{\meter^{-2}}$, $\beta = 1.25 \alpha$, and $\xi=0.5$ (right). Dashed gray line shows $\tilde{\rho}_t$. Inset of right panel shows the change $\Delta \tilde{\rho}_{\mathrm{tot}} / \tilde{\rho}_{\mathrm{tot,vel}}$, where $\Delta \tilde{\rho}_{\mathrm{tot}} = \tilde{\rho}_{\mathrm{tot,vel}} - \tilde{\rho}_{\mathrm{tot}}$ and $\tilde{\rho}_{\mathrm{tot,vel}}$ is the total cell density for the variation of the continuum model that assumes vertical cell transport, versus shifted radial coordinate $\tilde{r}$.}
\end{figure}

\newpage

\section*{Supplementary Figures 12-13: Fitting the continuum model to the agent-based model}

\begin{figure}[H]
\centering
\includegraphics[width=0.8\columnwidth]{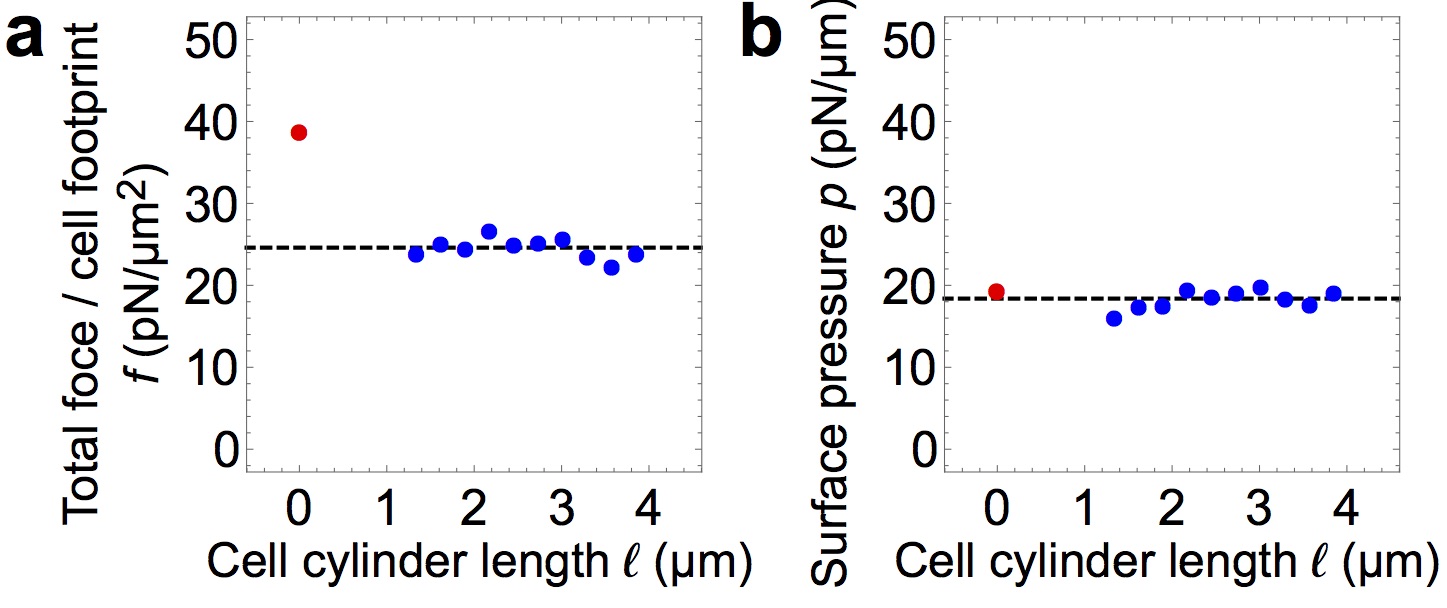}
\caption{\label{fig:FG9004}
Defining ``pressure'' in the agent-based model. (\textbf{a}) Average of normalized total force $f$, where we define $f$ as the total force on a cell divided by the projected area of the cell onto the surface (i.e., the cell footprint), and the average is taken over horizontal cells (blue data points) and vertical cells (red data point) in the mixed interior with projected cell cylinder length around $\ell$, versus projected cell cylinder length $\ell$ for ten simulated biofilms with initial cell cylinder length $\ell_0 = \SI{1.2}{\micrometer}$. Black dashed line shows average of $f$ over both horizontal and vertical cells. (\textbf{b}) Average surface pressure $p$, where we define $p$ as the total force on a cell divided by the perimeter of the cell footprint, and the average is taken over horizontal cells (blue data points) and vertical cells (red data point) in the mixed interior with projected cell cylinder length around $\ell$, versus projected cell cylinder length $\ell$ for ten simulated biofilms with initial cell cylinder length $\ell_0 = \SI{1.2}{\micrometer}$. Black dashed line shows the average of horizontal and vertical surface pressures.
}
\end{figure}

\textbf{Mapping the surface pressure in the continuum model to forces in the agent-based model} What is the relationship between the surface pressure in our continuum model and the microscopic, cell-cell contact forces? On the cell scale, the disorder of the cell configuration yields forces on cells that are extremely heterogeneous in space and time, even for a fixed radial distance along the moving front (Supplementary Fig. 10). A further contribution to this disorder comes from polydispersity in the cell lengths. That is, since larger cells can have more cell-cell contacts, we expect the forces acting on a cell to increase with cell size, on average. Therefore, to understand how the cell-cell contact forces relate to the surface pressure in the continuum model, we considered the forces acting on cells as a function of cell length.

For convenience, we focused on the mixed interior of the biofilm, since there our continuum theory predicts a uniform value of the macroscopic surface pressure. We found that the sum of the magnitudes of the in-plane forces on such cells scales with the perimeter of the cell footprint, but not with other quantities such as the cell footprint area (Supplementary Fig. 12), which is consistent with the behavior of an object embedded inside a two-dimensional, homogeneous fluid in mechanical equilibrium. Therefore, we quantified the surface pressure acting on a cell within the agent-based model as the sum of the magnitudes of the in-plane forces acting on a cell divided by the perimeter of its footprint. \\

\newpage

\begin{figure}[H]
\centering
\includegraphics[width=0.8\columnwidth]{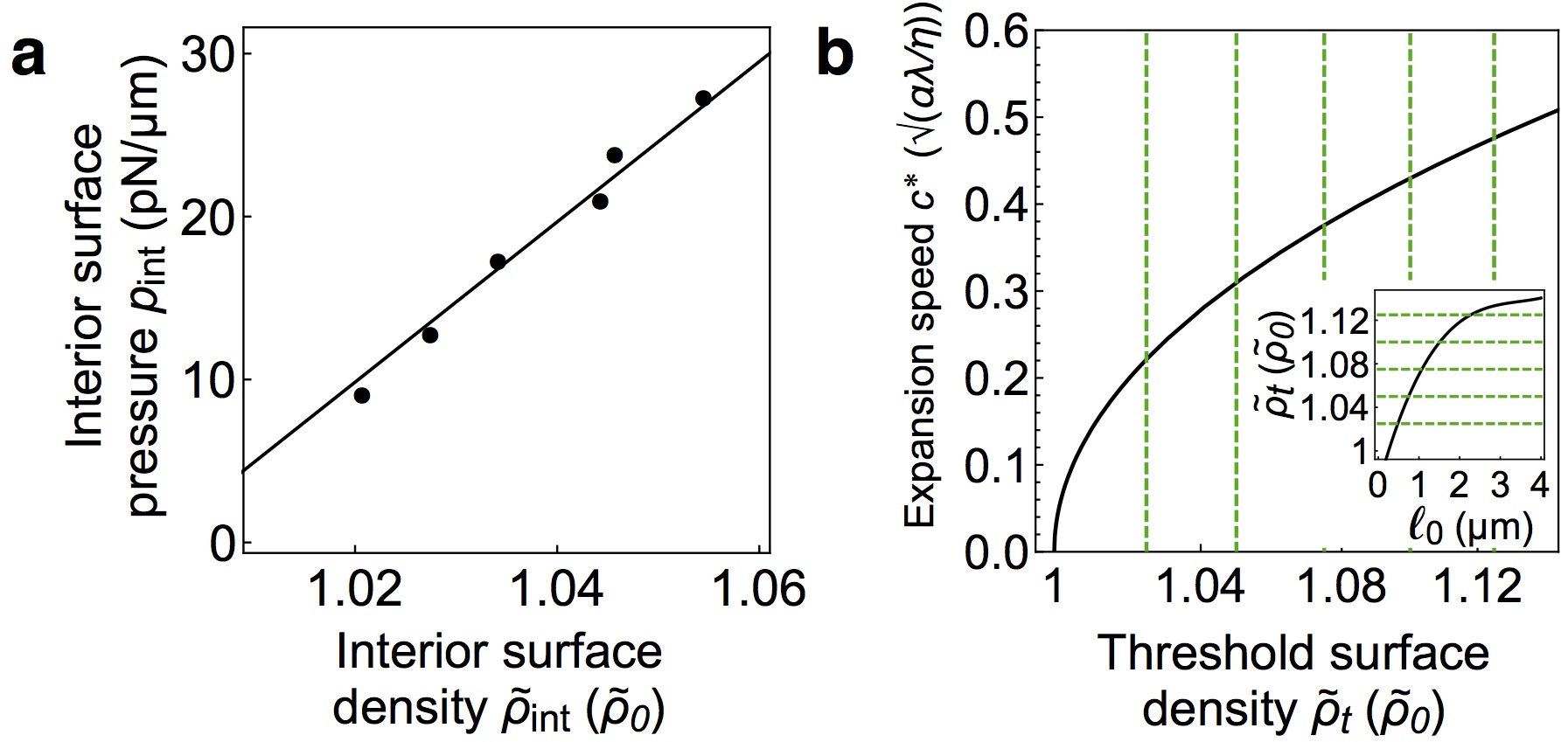}
\caption{\label{fig:FG9006}
Obtaining parameters for the continuum model from the agent-based model. (\textbf{a}) Average surface pressure $p_{\mathrm{int}}$ acting on cells in the mixed interior versus average interior density $\tilde{\rho}_{\mathrm{int}}$, where $\tilde{\rho}_{\mathrm{int}}$ is defined as the total footprint of cells in the mixed interior divided by the surface area of the mixed interior. Each data point is extracted using a different value of the initial cell length $\ell_0$, and averaged over ten simulated biofilms. (\textbf{b}) Interpolated expansion speed $c^*$ versus threshold surface density $\tilde{\rho}_{t}$, defined as the cell density in an annular window of $\SI{2}{\micrometer}$ centered at the radius of the maximum horizontal cell density, for agent-based models simulated over a range of different initial cell lengths. Inset shows the interpolated threshold surface density versus initial cell cylinder length. For reference, the dashed green lines in the main panel and the inset indicate the same five values of threshold surface density.
}
\end{figure}

\textbf{Choice of parameters for the continuum model} We fitted the parameters of our continuum model from results of the agent-based model as follows:

\begin{itemize}
\item { Cell stiffness $\lambda$: we fitted this parameter by measuring the average surface pressure and density in the central region of the biofilm for a range of initial cell lengths (Supplementary Fig. 13). A linear fit was then performed to extract $\tilde{\rho}_0$ and $\lambda$. Data were averaged over ten simulated biofilms. }
\item {  Threshold surface density $\tilde{\rho}_{t}$: the threshold surface density was calculated by averaging the pressure acting on all horizontal cells in a small radial window around the peak horizontal cell radial density (inset, Supplementary Fig. 13b). }
\item { The verticalization rate $\beta$ was obtained by fitting our continuum model for the horizontal cell density profile to the mixed interior of the biofilm in agent-based simulations, which has a decay constant given by Supplementary Eq. \ref{eq:dec}. For $\ell_0 = 1$, comparable to those in our experiments, we find $\beta=\SI{2.5}{hour^{-1}}$.}
\item { The ratio of vertical to horizontal cell footprints $\xi$: we computed $\xi$ as the average ratio of footprints in the mixed interior, where we defined footprint as the projected area of the cell onto the surface. }
\item { The cell growth rate $\alpha$ was determined from Supplementary Eq. \ref{eq:td} as $\alpha = \log 2 / t_d$, which yields $\alpha = \SI{1.4}{hour^{-1}}$.}
\item { The cell stiffness $\lambda$: we chose the cell stiffness equal to the Young's modulus times the cell radius, with the Young's modulus $Y=\SI{450}{\pascal}$ measured from bulk rheology, which yields $\lambda=\SI{360}{\pico\newton/\micrometer}$.}
\item { Surface viscosity coefficient $\eta$: we determined $\eta$ by fitting the expansion speeds in the continuum model to those of the agent-based model (Fig. 4a), which yields $\eta \simeq 10^5\ \SI{}{\pascal\second}$. }
\end{itemize}

\newpage

\section*{\textbf{Supplementary Figure 14: The effect of surface curvature}}

\begin{figure}[H]
\centering
\includegraphics[width=0.4\columnwidth]{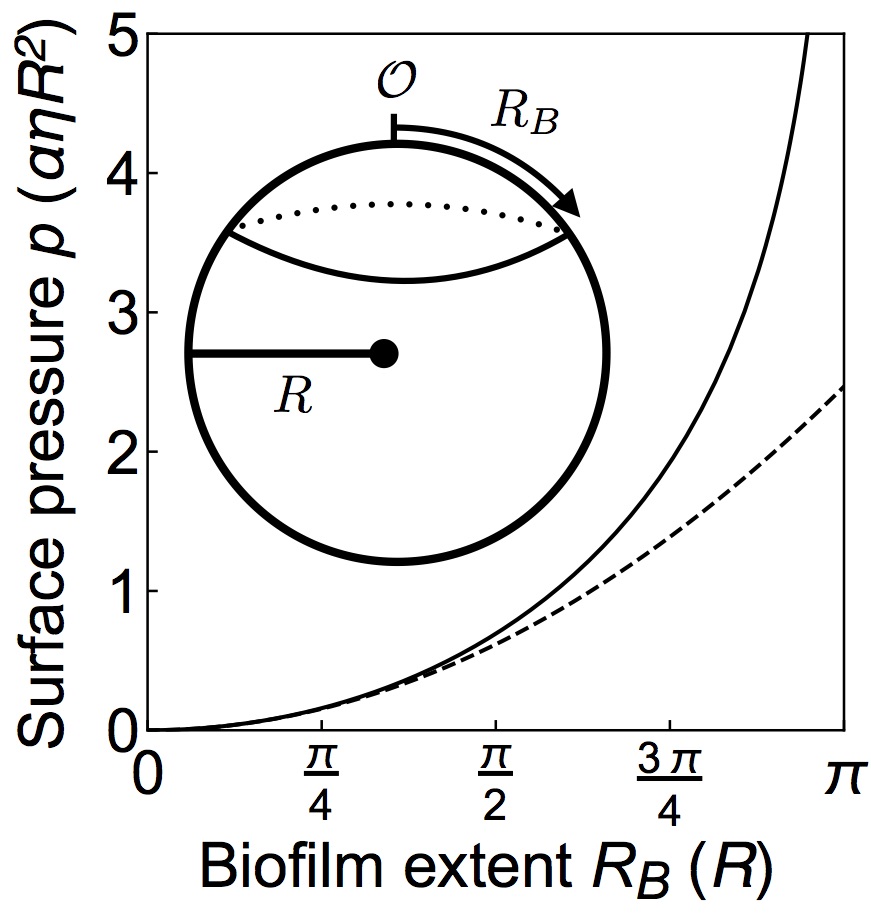}
\caption{\label{fig:FG9007}
Surface pressure $p$ at the origin $\mathcal{O}$ of growth versus biofilm extent $R_B$ for expansion along the surface of a sphere (solid curve) and a flat surface (dashed curve). Inset shows schematic illustration of biofilm expansion along the surface of a sphere.}
\end{figure}

\newpage

\section*{Supplementary Figure 15: Analysis of biofilms with cell-to-cell adhesion}

\begin{figure}[H]
\centering
\includegraphics[width=0.8\columnwidth]{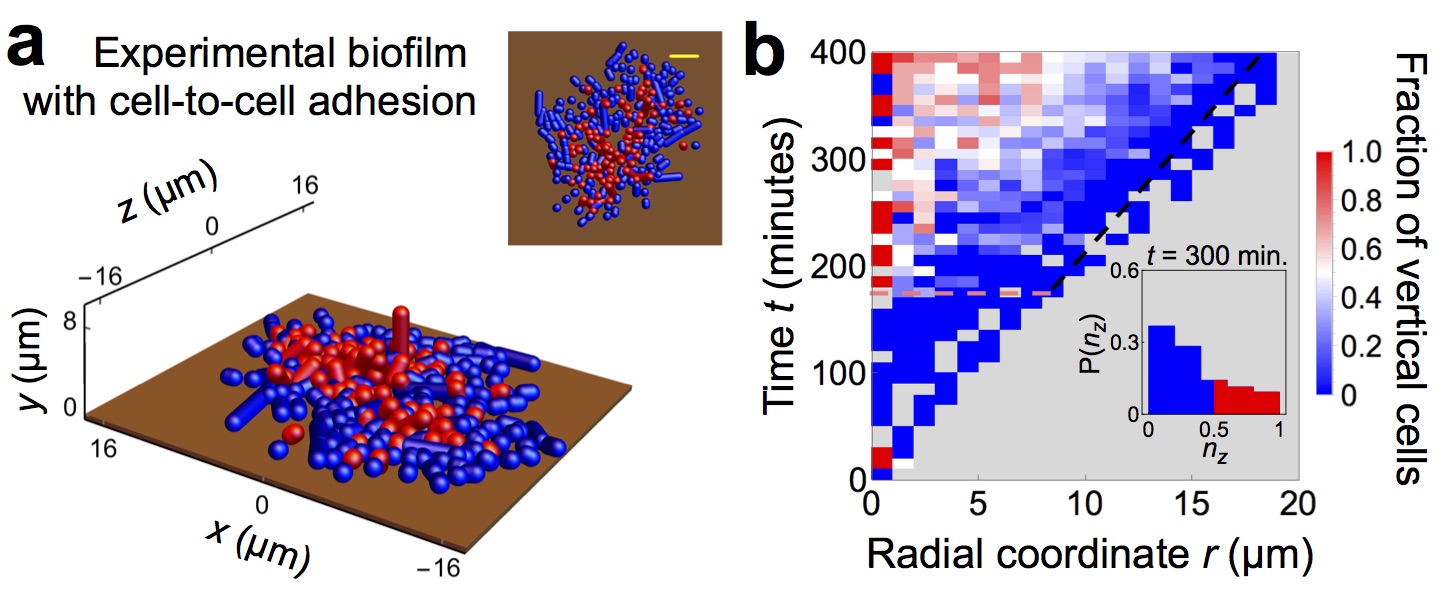}
\caption{\label{fig:FG9001}
Development of experimental biofilms with cell-to-cell adhesion. (\textbf{a}) Top-down and perspective visualizations of the surface layer of a living biofilm with cells producing the cell-cell adhesion protein RbmA, showing positions and orientations of horizontal (blue) and vertical (red) surface-adhered cells as spherocylinders of radius $R=\SI{0.8}{\micrometer}$, with the surface shown at height $z=\SI{0}{\micrometer}$ (brown). Cells with $n_z < 0.5$ ($>0.5$) are considered horizontal (vertical), where $\boldsymbol{\hat{n}}$ is the orientation vector. Scale bar: $\SI{5}{\micrometer}$. (\textbf{b}) 2D growth of a biofilm surface layer containing cells that produce RbmA. The color of each spatiotemporal bin indicates the fraction of vertical cells at a given radius from the biofilm center, averaged over the angular coordinate of the biofilm (gray regions contain no cells). The dashed pink line shows the onset of verticalization. The black dashed line shows the edge of the biofilm. The insets shows the distribution of cell orientations at time $t=300$ minutes, with color highlighting horizontal and vertical orientations.
}
\end{figure}

For the majority of the data presented, we used a \emph{V. cholerae} strain in which the gene (\emph{rbmA}) encoding the cell-to-cell adhesion protein RbmA, was deleted. We also performed experiments with a \emph{V. cholerae} strain in which the \emph{rbmA} gene was present and so RbmA protein was produced at wild type levels (Supplementary Fig. 15). We found that the transition to verticalization still occurred at roughly the same time, albeit with somewhat reduced vertical ordering. Interestingly, the horizontal expansion speed of the RbmA$^+$ biofilm was roughly $20\%$ more rapid than the biofilm we considered in the main text. This increase in expansion speed is consistent with the reduction in vertical ordering that we observed for the RbmA$^+$ biofilm.

\newpage

\section*{Captions for Supplementary Videos}

\textbf{Supplementary Video 1} Growth of a \emph{V. cholerae} biofilm cluster, showing cross-sectional images of the bottom cell layer. The strain constitutively expresses \emph{mKO}. The viewing window is 45 by 45 $\SI{}{\micrometer^2}$ and the total duration is 8 hours with 10 min time steps.

\textbf{Supplementary Video 2} Visualization of the surface layer of a modeled biofilm with initial cell cylinder length $\ell_0 = \SI{1}{\micrometer}$, showing positions and orientations of horizontal (blue) and vertical (red) surface-adhered cells as spherocylinders of radius $R=\SI{0.8}{\micrometer}$, with the surface shown at height $z=\SI{0}{\micrometer}$ (brown). Cells with $n_z < 0.5$ ($>0.5$) are considered horizontal (vertical), where $\boldsymbol{\hat{n}}$ is the orientation vector. Scale bar: $\SI{3}{\micrometer}$. The total duration is 10 hours.

\textbf{Supplementary Video 3} Visualization of the surface layer of a modeled biofilm with initial cell cylinder length $\ell_0 = \SI{2}{\micrometer}$, showing horizontal (blue) and vertical (red) cells as spherocylinders, the surface (brown), and cell-to-cell contact forces (yellow lines connecting the centers of cells, with thicknesses proportional to the force). Cells with $n_z < 0.5$ ($>0.5$) are considered horizontal (vertical), where $\boldsymbol{\hat{n}}$ is the orientation vector. The length of the scale bar is $\SI{3}{\micrometer}$, and its thickness corresponds to $\SI{300}{\pico\newton}$.

\textbf{Supplementary Video 4} Numerical simulation of the continuum model assuming no vertical cell transport, showing radial densities of horizontal cells ($\rho_h$, blue), vertical cells ($\rho_v$, red), and total density ($\tilde{\rho}_{\mathrm{tot}}=\rho_h + \xi \rho_v$, black), versus radial coordinate $r$ for isobaric regime with $\tilde{\rho}_0 = \SI{1}{\meter^{-2}}$, $\tilde{\rho}_t = \SI{1.5}{\meter^{-2}}$, $\beta = 2.5 \alpha$, and $\xi=0.5$. Dashed gray line shows $\tilde{\rho}_t$.

\textbf{Supplementary Video 5} Expansion of \emph{V. cholerae} biofilm clusters grown with the drug A22 at a concentration of $\SI{0.4}{\microgram/\milli\liter}$ (left) and the drug Cefalexin at a concentration of $\SI{4}{\microgram/\milli\liter}$ (right). Cross-sectional images of the bottom cell layers are shown. The strain constitutively expresses \emph{mKO}. Scale bar: $\SI{30}{\micrometer}$. The total duration is 10 hours with 30 min time steps.

\newpage

\section*{\textbf{Supplementary Discussion}}

\subsection*{\textbf{The effects of cell-scale geometrical and mechanical properties}} We have focused the current analysis on mutant \emph{V. cholerae} biofilms that have been engineered to have simpler interactions than wild type cells in biofilms (see Methods). For example, wild type cells are slightly curved\cite{S7}, and produce a cell-to-cell adhesion protein called RbmA\cite{S3}. The presence of cell curvature endows the cell with an additional rotational degree of freedom, which provides an additional direction along which a mechanical instability can proceed. In curved cells, due to the reduced extent of the cell in the transverse direction, we expect such an instability to occur with a lower reorientation threshold. This effect could allow verticalization to occur at lower values of surface pressure by proceeding in two stages: first by rotating away from the surface along the transverse direction, and then by being peeled off the surface at the remaining point(s) of contact. To understand how cell-to-cell adhesion could influence verticalization, we preliminarily analyzed biofilms that produce the cell-to-cell adhesion protein and found that the horizontal to vertical transition still occurs, albeit with somewhat reduced vertical ordering (Supplementary Fig. 15). Understanding the modifying effects of cell curvature and cell-cell adhesion will be important directions for future studies.

Our agent-based model does not explicitly incorporate the VPS matrix secreted by cells. A more detailed treatment of the VPS matrix could potentially explain small differences we observed between the orientational patterning and spreading dynamics of the agent-based model biofilms and those of the experimental biofilms. Furthermore, previous studies have shown that matrix production can allow the bacterial cells of the biofilm to establish an osmotic pressure difference between the biofilm and the external environment\cite{S9}, which could potentially impact the mechanics of the verticalization transition. Thus, understanding the interplay between cell and matrix mechanics will be an important direction for future studies.

By varying the parameters in our agent-based model that reflect cell-scale features, we observed a wide range of biofilm architectures of varying size, shape, orientational ordering, and dimensionality. Importantly, with regard to the features we analyzed, we found that the verticalization transition relies primarily on the presence of cell-to-surface adhesion, and so we expect our results to apply to a wide range of bacterial biofilms. In particular, our findings on mechanical instabilities are general enough to describe analogous transitions for other cell shapes, including spherical cells \cite{S10, S11, S12}, for which compression will induce vertical center of mass displacements. There are other types of biofilm architectures that we did not observe in our simulations. For example, \emph{Bacillus subtilis} have been observed to create planar Y-shaped formations, which appear to have an extended bending mode\cite{S13}. In addition, \emph{Escherichia coli} colonies that are compressed against the surface undergo a variant of the 2D-3D transition\cite{S14}, but with the cells reportedly remaining horizontal in a layered, wedding-cake type formation\cite{S15}. \emph{Pseudomonas aeruginosa} biofilms can form 3D streamers under the influence of flow\cite{S16}. These examples of known architectures already suggest a grand challenge in the study of biofilms: we must develop a systematic method to account for the diversity of architectures that can be produced by local mechanical interactions.

\subsection*{\textbf{The effects of surface curvature}}  For simplicity, we have considered expansion of biofilms along a flat surface. However, many surfaces in nature are curved, which would locally change the cell-to-cell and cell-to-surface contact geometries, as well as globally influencing the build-up of pressure throughout a biofilm. We expect the resulting changes in cell-cell contact geometry to decrease the threshold surface pressure for reorientation by facilitating the application of verticalizing torque. On the other hand, changes in the cell-to-surface geometry can both increase and decrease the threshold surface pressure for verticalization depending on the sign of the curvature. Thus, cells on a concave surface might undergo more spreading in two dimensions, while those on a convex surface might undergo more rapid three-dimensional expansion. Finally, surface curvature can increase or decrease the rate at which pressure builds up throughout a biofilm, since spreading out to a given distance has the consequence of covering a smaller or larger footprint depending on whether the surface is ball-like or saddle-like, respectively. This effect becomes significant when the extent of the biofilm along the surface becomes comparable to the radius of surface curvature (Supplementary Note), and thus could serve as an additional mechanism to modulate the onset of vertical growth.

\subsection*{\textbf{Evolution and adaptation of global biofilm morphology}} Our results suggest that bacteria have harnessed the physics of mechanical instabilities to generate complex architectures. What impact does a biofilm's morphology have on its growth and survival? \emph{V. cholerae} biofilm clusters have been observed to form as monocultures that exclude competitors\cite{S17}. When two clusters impinge upon one another, for example during resource competition, the structural properties of a biofilm become crucial determinants of its success in edging out competitors\cite{S17}. The morphology of a biofilm could also be important in driving how biofilm cells access nutrients. Nutrients can be delivered from surfaces, e.g. when the biofilm forms on a solid food source such as chitin or marine snow\cite{S18}, as well as by the surrounding fluid, e.g. via the diffusion of oxygen and other chemicals. Therefore, since both two-dimensional and three-dimensional growth can be beneficial, we expect a balance between horizontal and vertical growth to be most advantageous. We therefore speculate that individual cell features have evolved in response to selective pressures on the global morphologies of biofilm communities\cite{S10, S11}. Since optimal morphology may be condition dependent, cells may also have evolved adaptive strategies for biofilm formation, which could be investigated experimentally by screening environmental influences on cell size, shape, and surface adhesion. Intriguingly, data exists which suggests that, in nature, \emph{V. cholerae} undergoes morphological changes during starvation, including developing into small cocci and long filaments\cite{S12}.

\subsection*{\textbf{Dynamical isobaricity}} Our study of a two-fluid model for verticalizing biofilms led us to discover a novel type of front propagation in which mechanical feedback stabilizes a linearly-expanding density profile. Remarkably, this density profile is precisely uniform in the biofilm interior starting at some finite distance from the front, whereas previous models of front propagation saturate continuously toward uniformity \cite{S22, S23, S24}. The spatial uniformity is a hallmark of an isolated fluid in mechanical equilibrium. However, in our system, the internal state and volume of the biofilm surface layer are constantly changing due to cell growth. Verticalizing biofilms thus provide a striking example of how equilibrium-like features can emerge naturally in a system that is driven far from equilibrium. Indeed, the self-organized nature of this process yields a universal behavior for the expansion speed that is independent of details of the mechanical feedback, including the verticalization rate $\beta$ (which sets the rate of feedback) and the ratio of vertical to horizontal footprints $\xi$ (which sets the strength of feedback).

\subsection*{\textbf{Fluctuations in biofilm shape and pressure}} Our continuum model, along with our agent-based model, both address the case of growth in nutrient-rich conditions, which pertains to our experiments. These models capture the observations of compact, circular morphologies. However, bacterial range expansions have also been studied under nutrient-poor conditions, which are known to cause branching morphologies\cite{S25}. Thus, it would be interesting to investigate the connections between mechanical and chemical feedback on biofilm growth.